\documentclass[fleqn,usenatbib,useAMS]{mnras}

\usepackage{hyperref}
\usepackage{mathptmx}

\usepackage[T1]{fontenc}
\usepackage{ae,aecompl}
\pdfoutput=1

\usepackage{graphicx}	
\usepackage{amsmath}	
\usepackage{amssymb}	
\usepackage{lscape}
\usepackage{ulem}

\newcommand{\kms}{km~s$^{-1}$}

\newcommand{\msun}{M_{\odot}}

\title[Measuring the influence of the closest companion]
{Galaxy pairs in the Sloan Digital Sky Survey - XI. A new method for measuring the influence of the closest companion out to wide separations}

\author[D. R. Patton et al.]{
David R. Patton$^{1}$\thanks{E-mail: dpatton@trentu.ca}
Farid D. Qamar,$^{1,2}$
Sara L. Ellison,$^{3}$
Asa F. L. Bluck,$^{3,4}$
Luc \newauthor Simard,$^{5}$ 
J. Trevor Mendel,$^{3,6}$
Jorge Moreno$^{7,8,9}$
and Paul Torrey$^{10,9}$
\\
$^{1}$Department of Physics and Astronomy, Trent University, 1600 West Bank Drive, Peterborough, ON K9L 0G2, Canada.\\
$^{2}$Department of Physics, Engineering Physics \& Astronomy, Queen's University, Kingston, ON K7L 3N6, Canada\\
$^{3}$Department of Physics and Astronomy, University of Victoria, Victoria, BC  V8P 1A1, Canada\\
$^{4}$Institute for Astronomy, ETH Zurich, 27 Wolfgang-Pauli-Strasse, Zurich, 8093, Switzerland\\
$^{5}$National Research Council of Canada, 5071 West Saanich Road, Victoria, BC V9E 2E7, Canada\\
$^{6}$Max-Planck-Institut f\"ur extraterrestrische Physik, D-85748 Garching, Germany\\
$^{7}$Department of Physics and Astronomy, California State Polytechnic University, Pomona, CA 91768, USA\\
$^{8}$TAPIR, Mailcode 350-17, California Institute of Technology, Pasadena, CA 91125, USA\\
$^{9}$Harvard-Smithsonian Center for Astrophysics, 60 Garden Street, Cambridge, MA, 02138, USA\\
$^{10}$MIT Kavli Institute for Astrophysics \& Space Research, Cambridge, MA, 02139, USA\\
}

\pubyear{2016}

\begin{document}
\label{firstpage}
\pagerange{\pageref{firstpage}--\pageref{lastpage}}
\maketitle

\begin{abstract}

We describe a statistical approach for measuring the influence that a galaxy's closest companion has 
on the galaxy's properties out to arbitrarily wide separations.  We begin by identifying the closest companion for every galaxy in a large 
spectroscopic sample of Sloan Digital Sky Survey galaxies.  We then characterize the local environment of 
each galaxy by using the number of galaxies within 2 Mpc and by determining the isolation of the galaxy pair 
from other neighbouring galaxies.  We introduce a sophisticated algorithm for creating a statistical control sample 
for each galaxy, matching on stellar mass, redshift, local density and isolation.  
Unlike traditional studies of close galaxy pairs, this approach is effective in a wide 
range of environments, regardless of how far away the closest companion is (although a very distant 
closest companion is unlikely to have a measurable influence on the galaxy in question).
We apply this methodology to measurements of galaxy asymmetry, and find  
that the presence of nearby companions drives a clear enhancement in galaxy asymmetries.  
The asymmetry excess peaks at the smallest projected separations ($< 10$ kpc), where the mean asymmetry is enhanced 
by a factor of $2.0 \pm 0.2$.
Enhancements in mean asymmetry decline as pair separation increases, 
but remain statistically significant (1-2$\sigma$) out to projected separations of at least 50 kpc.

\end{abstract}




\section{Introduction}

The diverse population of galaxies seen in our local universe is the end product of 
hierarchical galaxy formation and the combined influence of various 
evolutionary processes.   While galaxy-galaxy mergers contribute to the 
growth of galaxies over time, interactions and mergers can also alter and 
sometimes transform the properties of galaxies in the process, 
thereby driving the evolution of many galaxies. 
Much of the observational evidence for these effects is derived from 
studies of galaxies which exhibit strong morphological disturbances 
and/or galaxies which have a close companion.  
It is now well established that, on average, these galaxies have 
enhanced star formation rates \citep{barton00,ellison08a,woods10,scudder12,patton13,davies15},
higher asymmetries \citep{hernandeztoledo05,patton05,hernandeztoledo06,depropris07,plauchu10a,plauchu10b,ellison10,casteels14},
lower nuclear metallicities \citep{kewley06,ellison08a,micheldansac08,kewley10,scudder12,ellison13} 
and increased AGN activity \citep{alonso07,ellison11,silverman11,liu12,ellison13,khabiboulline14,satyapal14} 
compared with relatively isolated and/or undisturbed galaxies.  

The changes seen in the observed properties of interacting galaxies are consistent with predictions from high resolution merger simulations, 
which show that strong gravitational interactions can drive low metallicity gas to the central 
regions of galaxies, triggering intense star formation and fuelling AGN activity 
\citep{mihos96,springel05,dimatteo07,cox08,montuori10,torrey12,scudder12,hopkins13,patton13,moreno15,scudder15}.  
Moreover, these simulations predict that the effects of these interactions may 
persist long after a close encounter or merger has taken place.
In the case of galaxy pairs, this implies that these effects may be present 
even when the galaxies no longer qualify as close pairs.
For example, the simulations of \citet{patton13} indicate that star formation can remain elevated for more than 
one Gyr after the first peri-centre passage, by which time the galaxies may be 100-200 kpc apart from one another.
These predictions suggest that observational studies need to move beyond 
close pairs and strongly disturbed systems for a full accounting of the 
effects of interactions and mergers on galaxy properties.

It is challenging to identify and interpret post-merger systems in situations where 
the merger did not occur relatively recently \citep{lotz08,ji14}.  However, in the case of 
galaxy pairs, it is in principle possible to identify interacting galaxies well after they have 
experienced close encounters.  Such systems may be seen at relatively close separations, 
particularly if they are on the verge of a subsequent close passage and imminent merger.  
On average, however, systems which are detected well after a close encounter will have relatively 
large projected separations \citep{patton13}, especially in the case of flyby interactions \citep{dimatteo07,sinha12}.   
By extending close pair studies out to 
wider pair separations, it should therefore be possible to obtain a more complete 
measure of the cumulative effects of interactions on galaxy properties.

There are already a number of galaxy pair studies which  
report differences in galaxy properties at pair separations of 
50-100 kpc \citep{ellison11,patton11,scudder12,casteels13,ellison13,khabiboulline14,satyapal14} 
and some which report differences beyond 100 kpc \citep{park07,li08,robaina09,park09,koss10,patton13}.
However, there are a number of challenges that arise when attempting to extend 
existing techniques out to such wide separations.  
First, any interaction-induced differences in galaxy properties are likely to be smaller in magnitude at larger separations, since 
these galaxies will have had (on average) more time to settle down since their 
most recent close encounter.  This effect will make it harder to distinguish the properties of these galaxies 
from those of their non-interacting counterparts, especially with small sample sizes.
As projected separation increases, the likelihood that a given companion will be 
physically associated also decreases, due to projection effects \citep{alonso04,nikolic04,edwards12}.  
At larger separations, it also becomes increasingly important to consider 
the competing influences of other neighbouring galaxies \citep{moreno13,karman15}.
Finally, larger scale environmental influences may become comparable to 
or more important than the influence of the closest companion for wider pairs
\citep{park07,moreno13,sabater13}.  

A number of approaches have been employed to address some of these issues.  
\citet{barton07} restrict their analysis to pairs which are relatively isolated from their 
surroundings.  Various studies have investigated the properties of galaxy pairs 
as a function of environment \citep{alonso04,mcintosh08,ellison10,lin10}, 
helping to separate the influences of interactions from those of larger scale environment.
\citet{robaina12} use a mock redshift catalogue to demonstrate how a correlation function approach 
to galaxy properties can yield biased results as pair separation increases.
Ultimately, a fundamental limitation of many pair studies arises when a single 
cut in projected separation is used to separate paired galaxies from relatively isolated 
galaxies which are used as a control sample.
While this approach is robust for relatively close pairs, 
it begins to break down at larger separations, since control galaxies will 
be restricted to progressively sparser environments, while some paired 
galaxies may have multiple close companions within the chosen threshold 
in projected separation. 

With this study, we aim to address these issues by introducing a new approach for 
classifying galaxies both in terms of their closest companions and 
their larger--scale environment.   This technique is specifically designed 
to be effective at detecting the influence of the closest companion out to 
wide projected separations, while being largely free of bias due to projection effects 
and other environmental influences.  An earlier version of this technique and data set 
was introduced briefly by \citet{patton13}, and has been used in several subsequent publications 
\citep{ellison15a,ellison15b,stierwalt15}.
This paper provides a comprehensive description of our methodology, which includes a sophisticated 
algorithm for creating statistical control samples.  

As a demonstration of this approach, we apply our methodology to measurements of
asymmetry for Sloan Digital Sky Survey (SDSS) galaxies (\S~\ref{secasym}).  
While it is now well established that galaxies in close pairs have enhanced asymmetries, 
these studies have not yet established the full spatial extent where asymmetry enhancements persist.
\citet{depropris07} find an excess in the fraction of asymmetric galaxies within about 60 kpc (40 $h^{-1}$ kpc), 
though it is unclear if this quantity levels off at larger separations.  
\citet{casteels14} report no excess in mean asymmetry beyond the relative projected separation 
of galaxy pairs (the separations at which the galaxy radii overlap) or beyond a projected 
separation of 35 kpc.  \citet{ellison10} find tentative evidence 
of an enhancement in the fraction of asymmetric galaxies that extends 
out to 80 kpc (the maximum $r_p$ of their pair sample), but find no clear convergence 
with their control sample at these large separations.   Our aim is to improve on these earlier results by comparing 
galaxy asymmetries with well matched control samples out to sufficiently large separations that 
the influence of the closest companion becomes negligible.

In the following section, we describe our data set and our metrics for identifying each galaxy's closest companion, 
along with our approach to characterizing each galaxy's environment.
In Section~\ref{seccontrol}, we outline the creation of statistical control samples for each galaxy.
We then address various sources of incompleteness in Section~{\ref{seccomplete}.
We investigate the influence of the closest companion on galaxy asymmetries in Section~\ref{secasym}. 
We end with our conclusions in Section~\ref{secconclusions}.
Throughout this paper, we assume a concordance cosmology of $\Omega_{\Lambda} = 0.7$, 
$\Omega_{\rm M} = 0.3$ and $H_0 = 70$ \kms Mpc$^{-1}$.

\section{Pair and Environmental Classification}\label{secsample}

\subsection{Methodology}\label{secmethods}

Our objective with this study is to systematically identify 
the closest companion for every galaxy in our sample, 
and to then detect the influence of each galaxy's closest companion 
using information about each galaxy's environment.
We wish to have an approach that is effective within a wide range of environments, 
from the low  density field to the cores of rich galaxy clusters.  Moreover, 
we would like our approach to be sensitive to the presence of 
other relatively nearby companions which may dominate over 
the influence of the closest companion.  
Finally, we would like to be able to detect the influence of the closest companion 
even if the companion lies at a relatively large separation and has had 
only a modest influence on the galaxy's current properties.

\subsection{Input Galaxy Catalogue}\label{secinput}

Following \citet{patton13}, we start by selecting all galaxies from the SDSS Data Release 7 \citep{dr7} 
which have reliable spectroscopic redshifts (zConf $>$ 0.7), 
along with extinction-corrected $r$-band Petrosian apparent magnitudes in the range of $14.0 \leq m_r \leq 17.77$.  
To avoid the extremes of the redshift distribution (which are more sparsely sampled), 
we also limit the redshift range to $0.005 < z < 0.2$.
We further require that every galaxy have a reliable total stellar mass estimate from 
\citet{mendel14}, which relies in part on the reprocessed SDSS photometry of \citet{simard11}.
We use the S\'ersic (rather than bulge plus disc) $ugriz$ total stellar mass fits of \citet{mendel14}, 
as recommended in their Appendix B.2.1.  These criteria yield a sample of 627 442 galaxies with 
secure estimates of redshift and stellar mass.

\subsection{Identifying Each Galaxy's Closest Companion}\label{secclose}

In this section, we outline our methodology for identifying the closest companion 
for every galaxy in the sample.  
We consider as potential companions only those galaxies which lie in our spectroscopic 
sample; however, we address potential sources of incompleteness in Section~\ref{seccomplete}.
We characterize potential companions using the most relevant available information 
on all galaxies in the vicinity: namely, projected physical separation from the galaxy in question (hereafter $r_p$), 
rest-frame relative velocity along the line of sight (hereafter $\Delta v$) and stellar mass.   

As our ultimate objective is to discern the influence of the closest companion 
on the properties of the galaxy in question, we  restrict our search for the closest companion to 
neighbouring galaxies which are sufficiently massive that they have the potential to exert a significant influence.  
In addition, we use $\Delta v$ primarily as a means to exclude unrelated foreground and background galaxies.
We define a potential companion to be any galaxy which 
has $\Delta v$ within 1000 km~s$^{-1}$ of the galaxy in question, and
which has a stellar mass which is at least 10 per cent of the stellar mass of the galaxy in question 
(i.e., a companion:host stellar mass ratio $\mu > 0.1$).
Of all potential companions meeting these two criteria, the galaxy with the smallest 
$r_p$ is deemed to be the closest companion.   
This general approach has been used in many earlier studies of close 
galaxy pairs \citep[e.g.,][]{barton00,patton00,lambas03,alonso04,ellison08a,patton11}.

For several reasons (including 
the environmental classifications described in \S~\ref{secenv} and the survey boundaries 
described in \S~\ref{secbound}), 
we cap $r_p$ at a maximum of 2 Mpc.  As a result, any galaxy which has no potential companions 
within 2 Mpc is deemed to have no detected closest companion, and is subsequently 
excluded from our analysis.  

Our relative velocity threshold of 1000 km~s$^{-1}$ is designed to be large enough to include 
the vast majority of companions which may conceivably be interacting with the 
galaxy in question, while minimizing contamination from unrelated foreground or background galaxies.   
All of the groups in the \citet{yang07} catalog (which includes both groups and clusters) 
have velocity dispersions < 1000 km~s$^{-1}$,  while more than 99 per cent of the 625 galaxy clusters 
in the catalogue of \citet{vonderlinden07} have velocity dispersions < 1000 km~s$^{-1}$, 
confirming that our threshold is a suitable choice.
However, we recognize that the likelihood of interaction increases 
as $\Delta v$ decreases.  Therefore, when studying candidate interacting systems, 
we restrict our analysis to galaxies which have a closest companion 
within 300 km~s$^{-1}$ (see \S~\ref{secsampleselect}).

Our minimum stellar mass ratio of $\mu = 0.1$ is intended to exclude potential companions which are of 
sufficiently low relative mass that they are unlikely to have had a significant influence 
on the galaxy in question.  Merger simulations suggest that this is a reasonable choice 
\citep{cox08,lotz10a}.   
We acknowledge that there will be cases where 
a nearby low mass companion is ignored despite exerting a greater influence than any more distant companions; 
however, given that the relative mass of any such companion is at most 10 per cent of the 
host galaxy's mass, this is likely to exclude relatively few cases in which the companion has had 
a meaningful influence on its more massive neighbour. 
Our stellar mass criterion does allow for the inclusion of galaxies 
which are close to a much more massive host galaxy (these systems will have high  
stellar mass ratios).  It is important to include these cases, since the massive companion 
will likely have a stronger influence than any other neighbouring galaxies.  However, 
as these cases will correspond to interactions between galaxies with very unequal masses (potential minor mergers), 
we will later focus on systems with more similar stellar masses 
by restricting our analysis to galaxies whose closest companions 
have a stellar mass ratio of $0.1 < \mu < 10$.

\subsection{Characterizing the Environment of Each Galaxy}\label{secenv}

In order to isolate the influence that a close companion has on a galaxy's properties, 
one must consider any competing effects from the galaxy's surrounding environment.
Various approaches to characterizing environment have been used in the literature, 
including membership and/or location within structures such as clusters, groups, sheets, filaments 
and voids \citep{yang07,darvish14,eardley15}, 
central/satellite classification \citep{kravtsov04}, 
local number density \citep{cooper05,baldry06}, 
clustering statistics \citep{robaina09,skibba13}, 
proximity to the nearest massive galaxy \citep{geha12,sanchezjanssen13,ruiz14,pearson16},  etc.

Given the particular goals of this study, and our desire to consistently classify a 
large number of galaxies within a flux-limited galaxy redshift survey, 
we elect to use two distinct metrics to describe each galaxy's environment.    
The first is the total number of detected companions\footnote{Following  
Section 2.3, we require all such companions to have $\Delta v < 1000$~\kms and 
$\mu > 0.1$ with respect to the galaxy being classified.} within 
a projected separation of 2 Mpc (hereafter $N_2$).  
This metric probes a scale which is roughly an order of magnitude larger than the separations 
within which earlier studies suggest that interactions have a measurable influence on galaxy properties 
\citep{park07,li08,robaina09,koss10}.  
$N_2$ is closely related to the projected number density of galaxies, which is a derived 
quantity that probes a similar length scale \citep[e.g.,][]{hogg03,baldry06,cebrian14}.  
However, $N_2$ is more straightforward to measure than number density, as it does not require 
one to correct for flux limits and spectroscopic incompleteness.  
We caution that the measured $N_2$ for a given galaxy will be influenced by 
the galaxy's stellar mass and redshift, since $N_2$ scales with the apparent 
number density of galaxies.  However, as we will use this quantity in 
a relative sense only (i.e. when creating matched control samples, as described in \S~\ref{seccontrol}), 
this lack of normalization has no bearing on our study.  

Our second metric is the projected distance to the galaxy's second closest 
companion (hereafter $r_2$).  
This parameter can be used to distinguish between pairs which are isolated from their surroundings (large $r_2$) 
versus those in more typical surroundings (moderate $r_2$) or in very crowded regions such as 
galaxy clusters or compact groups (small $r_2$).   We apply the same restrictions 
on $r_p$ ($< 2$ Mpc), $\Delta v$ ($< 1000$ km~s$^{-1}$) and stellar mass ratio ($> 0.1$)   
as we did when identifying each galaxy's closest companion (see \S~\ref{secclose}).  
As such, any galaxy with fewer than two companions within 2 Mpc will be of limited use in our 
analysis (though such galaxies may qualify as controls; see \S~\ref{seccontrol}).

While there will tend to be some correlation between 
$N_2$ and $r_2$, the fact that these two parameters are sensitive to 
different scales\footnote{While $r_2$ may lie anywhere in the allowed range of 0-2 Mpc,  
the clustering of galaxies skews this distribution to the smaller separation end of this range.}    
will allow us to distinguish between pairs in a broader range of environments than either parameter would 
permit on its own.  For example, at a fixed intermediate value of $N_2$, some pairs will be 
isolated from their surroundings (large $r_2$), while others may be strongly influenced by 
other nearby galaxies (e.g. if the pair is close to a massive host galaxy or lies within a compact group).  

\subsection{Examples of Pairs in Different Environments}

To illustrate the relationship between our chosen metrics and the 
variety of environments that they can probe, Fig.~\ref{figmockpair} 
depicts four hypothetical galaxies which each have a close companion 
(at a projected separation of 30 kpc) but which reside in different environments.  
Four different combinations of $N_2$ and $r_2$ are shown, with 
local density ($N_2$) increasing from the bottom row to the top row, 
and isolation ($r_2$) decreasing from the left column to the right column.
The galaxy being classified is at the centre of each panel (filled black symbol), 
with its closest and second closest companions depicted with blue and red symbols.
In the lower left hand panel of Fig.~\ref{figmockpair}, the galaxy and its 
closest companion are relatively isolated from all other surrounding galaxies (relatively large $r_2$) 
and reside in a low density environment on even larger scales (low $N_2$).  
In this case, the closest companion is likely to have a greater 
influence on the galaxy than other galaxies in its vicinity.  

\begin{figure}
\centerline{\rotatebox{0}{\resizebox{9.0cm}{!}
{\includegraphics{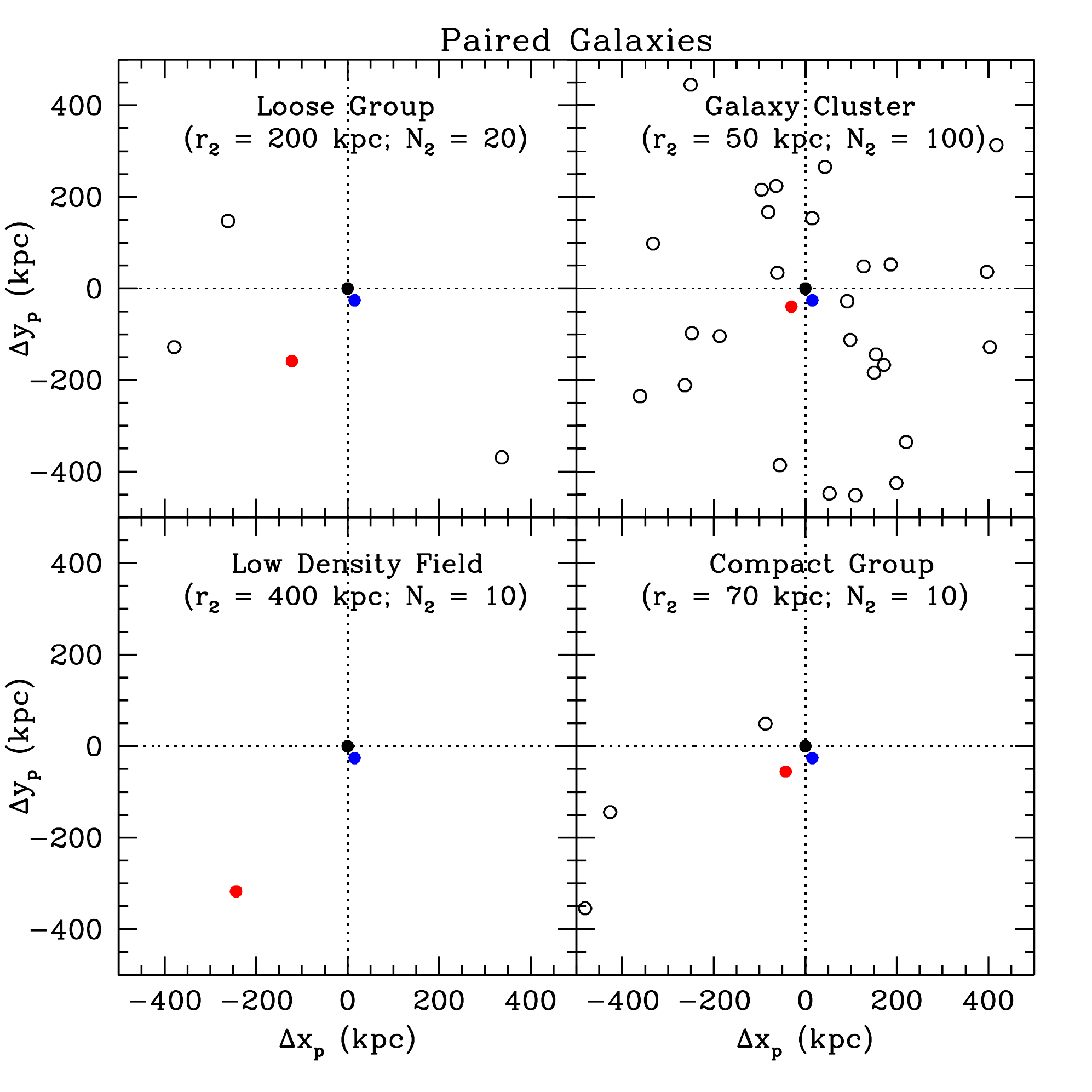}}}}
\caption{Hypothetical examples of close pairs are shown in four different environments, 
with local density ($N_2$) increasing from the bottom row to the top row, 
and isolation ($r_2$) decreasing from the left column to the right column.  
Each panel is 1000 kpc (projected) on a side (for clarity of presentation, we do not attempt 
to show all companions which lie within 2 Mpc of the paired galaxy).  
Within each panel, the galaxy which is being classified is located at the centre (filled black circle), 
the closest and second closest companions are shown with blue and red filled circles respectively, 
and all remaining galaxies are shown with open black circles.
\label{figmockpair}}
\end{figure}

In the remaining three cases, however, additional galaxies in the vicinity 
may have a more significant - and perhaps dominant - influence.
In the lower right panel of Fig.~\ref{figmockpair}, 
the galaxy pair resides within a compact group.
In this case, the galaxy resides in a relatively low density environment (low $N_2$), 
but the presence of several close companions (and corresponding low $r_2$) 
reduces the likelihood that the closest companion has a dominant influence on the galaxy in question.  
The upper left hand panel illustrates the more common scenario of the galaxy pair being in a loose 
group of galaxies, with intermediate values of both $r_2$ and $N_2$.
Finally, in the upper right panel of Fig.~\ref{figmockpair}, the pair resides within a 
galaxy cluster, with small $r_2$ and large $N_2$.  
In this case, the closest companion is one of many galaxies that lie 
fairly close to the galaxy in question, and it is quite possible that the galaxy's 
properties may be affected most by the cluster itself, 
rather than by the influence (if any) of its closest companion.  
By measuring both $N_2$ and $r_2$, 
we can discern between these different environments much better than we could 
with only one of these metrics.  Moreover, we will use these same metrics to 
identify effective control samples in different environments (\S~\ref{seccontrol}), 
which is essential for extending pair studies out to wider separations.

Having identified the closest companion for each of the galaxies in our SDSS sample, and having 
measured both $N_2$ and $r_2$ for each galaxy, 
it is now possible for us to select SDSS galaxy pairs which reside in a range of different environments.   
To illustrate this ability, we identify pairs which are similar to the hypothetical examples shown in Fig.~\ref{figmockpair}.
We select four pairs with separations of $r_p \sim 30$ kpc and with approximate matches in both $N_2$ and $r_2$.  
SDSS images of these representative pairs are displayed in Fig.~\ref{figanalogs}.
We note that the pair which lies in the ``Compact Group'' environment was identified as a compact 
group by \citet{mcconnachie09}.  

\begin{figure}
\centerline{\rotatebox{0}{\resizebox{9.0cm}{!}
{\includegraphics{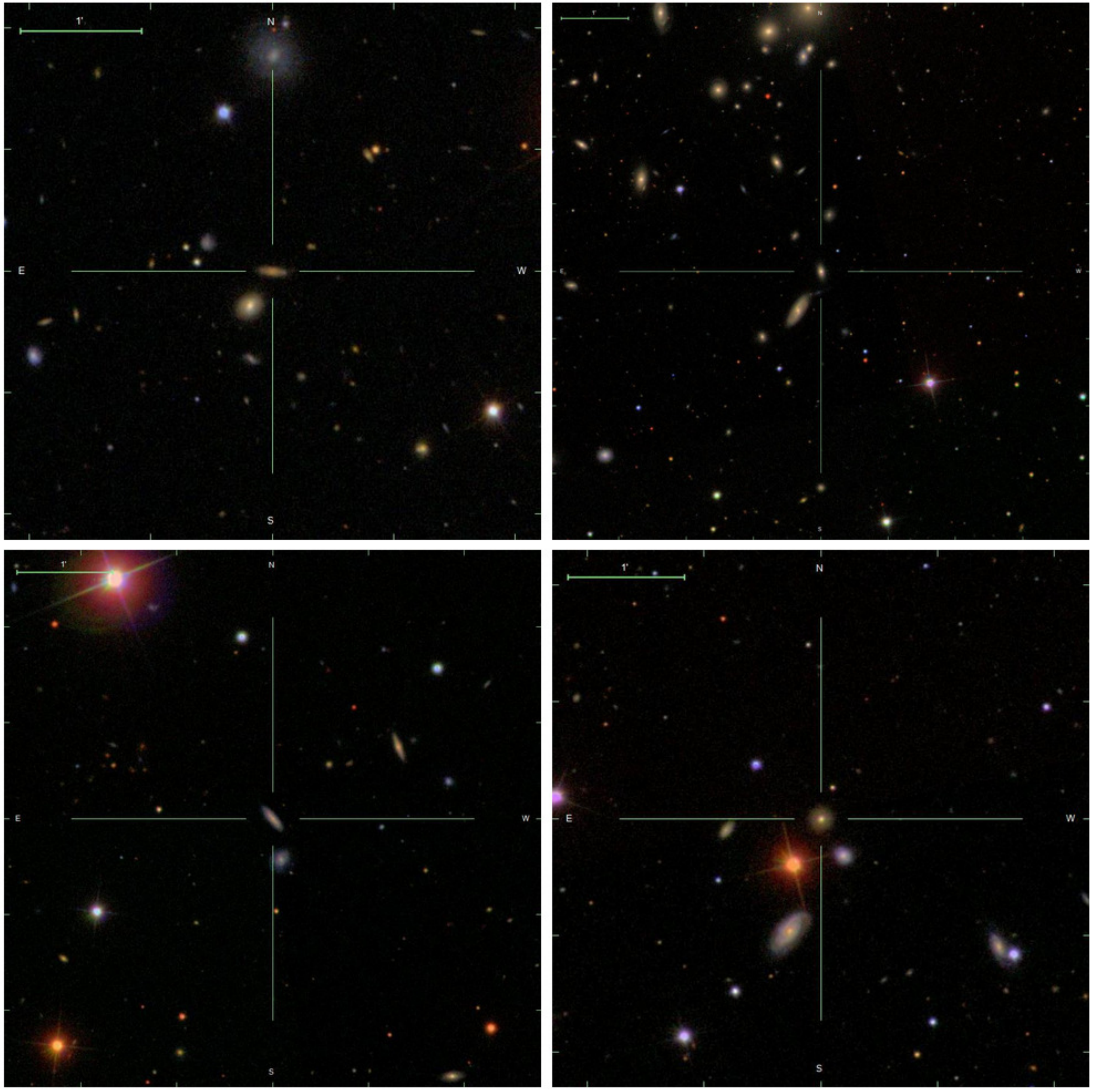}}}}
\caption{SDSS three-colour images are shown for four representative close pairs ($r_p \sim 30$ kpc) 
residing in different environments.  Each image is 200 kpc (projected) on a side, and is centred 
on one of the members of the close pair.  The green scale bar in the upper left of each 
image denotes an angular separation of one arcminute.
The four images are analogous to the corresponding four panels of Fig.~\ref{figmockpair}.  
The lower left panel shows an isolated pair in a low density environment ($N_2 = 11$; $r_2 = 471$ kpc), 
the lower right panel shows a pair in a compact group ($N_2 = 9$; $r_2 = 71$ kpc), 
the upper left panel shows a pair in a loose group environment ($N_2 = 22$; $r_2 = 189$ kpc),  
and the upper right panel shows a pair in a cluster-like environment ($N_2 = 104$; $r_2 = 44$ kpc).  
\label{figanalogs}}
\end{figure}

\section{Creation of Statistical Control Samples}\label{seccontrol}

\subsection{Methodology}

Having identified each galaxy's closest companion, and having also characterized the environment 
of each, we now address the challenge of detecting the influence (if any) of the closest companion 
on each galaxy.  To this end, we select a statistical control sample for each galaxy which is 
well matched in stellar mass, redshift and environment, with the only difference being that 
each control galaxy does not have a comparably close companion.  
By comparing galaxies with their controls, and averaging over many 
systems with similar properties (e.g. the distance to the nearest companion), 
we will demonstrate that clear differences can be detected between galaxies 
and their controls, with these differences being attributed to the presence and 
inferred influence of the closest companion.  

Our decision to match controls on stellar mass and redshift follows earlier galaxy pair studies 
\citep{ellison08a,perez09a,patton11,xu12}.  
Many galaxy properties depend on stellar mass, as illustrated by well established 
relationships such as the star formation rate - stellar mass main sequence \citep{noeske07,speagle14}, 
the mass-metallicity relation \citep{tremonti04,ellison08b}  
and the size-mass relation \citep{shen03,ichikawa12}.  
In addition, the spatial resolution, sample depth and survey volume 
vary with redshift, driving the main selection effects which are present in this redshift survey.  
By matching on stellar mass and redshift, we can ensure that any 
detected differences between paired galaxies and their controls 
are not driven by differences in these more fundamental properties.
Moreover, matching on these two properties also mitigates 
aperture effects when studying spectroscopic properties.

Some earlier galaxy pair studies have also matched control samples on environment, 
either by requiring paired galaxies and their controls to have similar environmental classifications 
\citep[e.g., group/cluster membership;][]{alonso04,alonso12} 
or by explicitly matching on local density \citep{perez09b,ellison10,scudder12,xu12}.
However, as these approaches typically characterize environment on scales substantially 
larger than the separations of the pairs themselves, 
they are unable to distinguish between the scenarios 
illustrated in Fig.~\ref{figmockpair}.

We instead create our control samples by simultaneously matching on 
both local density and isolation, as initially described by \citet{patton13}.
To match on local density, we use our measurements of $N_2$. 
Given that the galaxy and its controls are also matched 
in stellar mass and redshift, the search area included within 2 Mpc 
will be of a similar angular extent and photometric depth for a galaxy 
and its controls, making this a fair comparison and effectively a match 
on projected number density.  

To match on isolation, we require that the 
projected distance to each control galaxy's closest companion ($r_p$) be comparable 
to the projected separation of the {\it second closest companion} ($r_2$) of the galaxy in question.  
In other words, the isolation of the galaxy pair in question must be similar to the isolation 
of each of its individual control galaxies.
This matching is illustrated in Fig.~\ref{figmockpairc}, which depicts 
ideal control galaxies for the four hypothetical paired galaxies in Fig.~\ref{figmockpair}. 
Close comparison of these figures shows that the only differences 
between a paired galaxy and its ideal control is the presence of its closest companion; 
in all other respects, the surrounding environment is the same.  
In practice, of course, we cannot ensure a perfect match between the environments 
of any two galaxies.  However, by matching on both $N_2$ and $r_2$, 
we can find suitable controls for galaxies in a wide range of environments.

\begin{figure}
\centerline{\rotatebox{0}{\resizebox{9.0cm}{!}
{\includegraphics{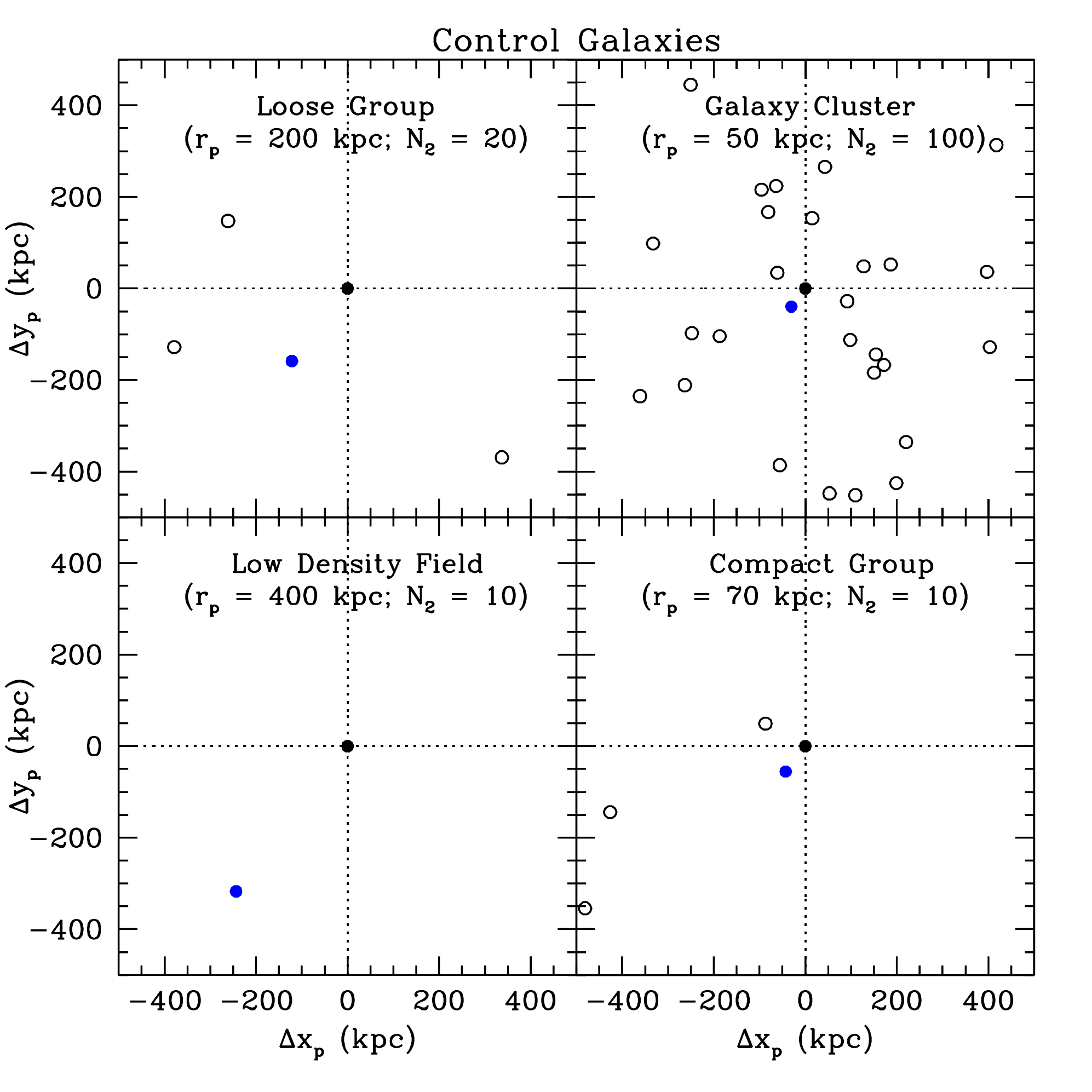}}}}
\caption{Ideal control galaxies and their surroundings are shown for 
the four paired galaxies depicted in Fig.~\ref{figmockpair}.
The scale and colour scheme is the same as in Fig.~\ref{figmockpair}, 
except for the fact that the second nearest companion is no longer explicitly 
identified (since it is irrelevant here).  In each case, the control galaxy's 
closest companion is the same as the paired galaxy's second closest companion 
in Fig.~\ref{figmockpair}.  As a result, the $r_p$ values in this figure are 
identical to the $r_2$ values in Fig.~\ref{figmockpair}.  
This figure illustrates how we match on both isolation and 
number density when creating our control samples.
\label{figmockpairc}}
\end{figure}

\subsection{Implementation}\label{secimplementation}

We now describe how we implement control sample matching on stellar mass, redshift, local density and isolation.  
Our objective is to maximize the number of control galaxies while simultaneously 
requiring good quality matches on all four quantities.
We begin by selecting a default matching tolerance for each of these quantities.  
For stellar mass, our default tolerance is 0.1 dex, which is comparable to the typical statistical uncertainties 
on the stellar mass measurements \citep{mendel14}.
We require the redshifts to agree to within 0.01, which is nearly two orders of magnitude 
larger than the measured uncertainties in SDSS DR7 spectroscopic redshifts \citep{dr7}.
To match in local density, we require the $N_2$ values to agree within 10\%.  
Finally, to match in isolation, we require that the $r_p$ of the control galaxy's closest companion be  
within 10 per cent of $r_2$ (the projected separation of the paired galaxy's second closest companion).  
We allow for replacement, meaning that a given galaxy may act as a control for more than one galaxy.  

For 84 per cent of galaxies in our sample, this approach yields at least ten control galaxies.  
For the remainder of the sample, we increase each of the tolerances by 50 per cent 
(yielding revised tolerances of 0.15 dex, 0.015, 15 per cent and 15 per cent) and repeat the process, 
continuing until at least ten controls are found for each galaxy.   
This procedure yields an average of 68 controls for each galaxy in our sample.

Given that some control galaxies are better matches than others, we then apply a weighting scheme which 
assigns larger statistical weights to better matches.  In particular, for each quantity that is being matched, we compute 
a statistical weight for each control galaxy, such that a perfect match will yield a weight of one, 
while the worst match (at the limits of the allowed tolerance) will yield a weight of zero.   
For example, for a galaxy with redshift $z$ and redshift tolerance $z_{\rm tol}$, 
the $i^{\rm th}$ control galaxy (with redshift $z_i$) is assigned a redshift weight of 
\begin{equation}\label{eqnwz}
w_{z_i} = 1 - {|z-z_i| \over z_{\rm tol}}.
\end{equation}
The overall statistical weight for a given control galaxy takes into account the quality of the match in all four 
quantities.  Continuing with the previous example, the overall statistical weight for the $i^{\rm th}$ control galaxy is 
given by
\begin{equation}\label{eqnw}
w_i = w_{z_i} w_{M_i} w_{N_2i} w_{r_2i},
\end{equation}
where the subscripts $M$, $N_2$ and $r_2$ refer to matches in stellar mass, $N_2$ and $r_2$ respectively.
Finally, these statistical weights can be used to compute the statistical mean of any desired quantity 
for a given galaxy's control sample.  For example, if a given galaxy has $N$ controls, the 
statistical mean of a given property $x$ of its statistical control sample is given by
\begin{equation}\label{eqnwc}
x_c = {\sum_{i=1}^N w_i x_i \over \sum_{i=1}^N w_i}.
\end{equation}

\subsection{Validation}\label{secvalidation}

In this section, we investigate the effectiveness of our control sample algorithm by assessing
the quality of the matches and the benefits of applying statistical weights.
In Fig.~\ref{figmethodshist}, we display histograms of the four quantities 
that are matched: redshift, stellar mass, $N_2$ and $r_2$.   We compare 
histograms for paired galaxies (blue symbols) and their weighted mean 
controls (red symbols).  We find very good agreement between 
the redshifts of galaxies and their statistical controls.
Agreements in stellar mass and $N_2$ appear to be even tighter, such that the controls are 
nearly indistinguishable from the galaxies they are matched to.  
The matching on $r_2$ is excellent at most separations, but diverges 
at the largest projected separations probed.    
This disagreement stems from the fact that $r_p$ and $r_2$ are 
not allowed to be greater than 2 Mpc (see \S~\ref{secenv}).  As a result, when considering 
a galaxy which has $r_2$ close to 2 Mpc, all potential controls will have $r_p < 2$ Mpc; 
this biases the control sample such that $r_p$ will tend to be smaller than $r_2$.  
This is precisely the behaviour that is seen in the 
upper right panel of Fig.~\ref{figmethodshist}, where controls scatter out of the 
range $1900 \lesssim r_2 < 2000$ kpc and into the range $1600 \lesssim r_2 \lesssim 1900$ kpc.
In order to avoid this issue, we subsequently restrict our analysis to galaxies which have $r_2 < 1500$ kpc.
This restriction has the additional effect of reducing the average number of controls per galaxy 
from 68 to 62.

\begin{figure}
\centerline{\rotatebox{0}{\resizebox{9.0cm}{!}
{\includegraphics{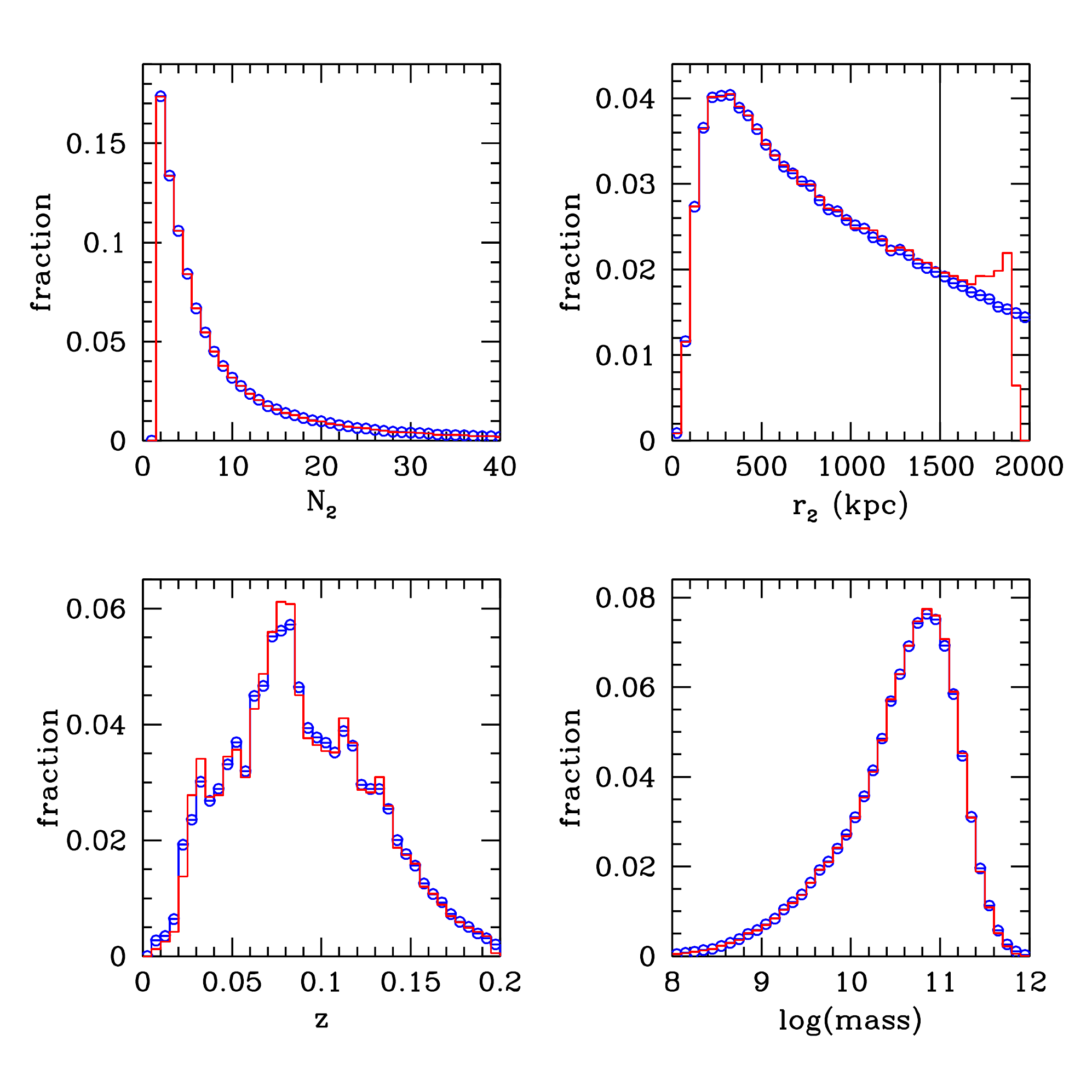}}}}
\caption{Histograms of redshift, stellar mass, $N_2$ and $r_2$ are shown for 
paired galaxies (blue squares).  The corresponding histograms for their weighted 
mean controls are depicted using red symbols.  However, in the upper right hand plot, 
the control sample histogram denotes $r_p$ (rather than $r_2$), since 
this is the quantity that is matched to the paired galaxy $r_2$ values.
All histograms are normalized such that the area under the histogram is equal to one.  
Overall, excellent agreement is seen between paired galaxies and their controls. 
The only regime in which substantial disagreement is seen is 
in the upper right hand panel, where $r_2 \gtrsim 1500$ kpc.  
As described in the text, we subsequently exclude from our analysis 
all paired galaxies with $r_2 > 1500$ kpc, as indicated with the vertical dashed line.
\label{figmethodshist}}
\end{figure}

While the overall distributions of redshift, stellar mass, $N_2$ and $r_2$ for paired galaxies 
and their controls appear to agree well for $r_2 < 1500$ kpc, a statistical comparison between galaxies 
and their controls is warranted.  
To compare redshifts, we compute $\Delta_z$, which we define as the average difference between a galaxy's 
redshift ($z$) and the mean redshift of its controls ($z_c$), such that 
\begin{equation}
\Delta_z = |<z-z_c>|,
\end{equation}
Analogous terms are computed for stellar mass, $N_2$ and $r_2$.  
These results are reported in Table~\ref{tab1}, for three different versions of the control samples.  
First, we use only the single best match for each galaxy (the control galaxy with the highest 
statistical weight), which is called the ``Best Match'' control sample.
We find very small differences for all four properties, with all of these differences being 
much smaller than the default tolerances used in our matching algorithm.  For example, the mean stellar mass of paired 
galaxies differs from the controls by only 0.0029 dex, which is substantially smaller than our default 
matching tolerance of 0.1 dex in stellar mass.  In fact, this difference is considerably smaller than 
the $\sim 0.1$ dex uncertainties on the stellar mass measurements themselves \citep{mendel14}, 
indicating that our matching algorithm is more than sufficient to remove any 
measurable difference between the stellar masses of paired galaxies and their controls.

Secondly, we use all suitable control galaxies, but assign an equal weight to each.  We report 
these results in the ``Unweighted Mean'' row of Table~\ref{tab1}.  
In this case, using a wider range of control galaxies leads to poorer matches, 
with $\Delta$ values increasing by factors of $\sim$ 3-5 
compared with the best match approach.  
However, as this approach uses an average of 62 control galaxies for each paired galaxy, 
it will provide a more uniform and representative control sample for each galaxy.

Finally, we use our preferred ``Weighted Mean'' control sample for each galaxy, 
applying the weighting scheme outlined in Section~\ref{secimplementation}.  
In this case, we use the same control galaxies as for the unweighted means, 
but the matches are now tighter by a factor of $\sim$ 1.7-2, 
based on the corresponding decrease in the $\Delta$ values in Table~\ref{tab1}.
In other words, by using our weighting scheme, we are able to increase the average size 
of our control sample by a factor of 62, while sacrificing only a factor of two in the 
quality of the control sample matching.

\begin{table}
\centering
\caption{Statistical Comparison of Paired Galaxies and Their Controls\label{tab1}}
\begin{tabular}{ccccc}
\hline
Control Sample&$\Delta_z$&$\Delta_{\rm mass}$&$\Delta_{N_2}$&$\Delta_{r_2}$\\
&&(dex)&&(kpc)\\
\hline
Best Match& 0.000018& 0.0029& 0.028& 0.78\\
Unweighted Mean& 0.000086& 0.0095& 0.097& 2.81\\
Weighted Mean& 0.000043& 0.0057& 0.055& 1.57\\
\hline
\end{tabular}
\end{table}

Having demonstrated overall agreement between galaxies and their controls, 
we now examine the effectiveness of our control sample matching as a function of $r_p$.  
In Fig.~\ref{figcontrend}, we plot the mean redshift, stellar mass, $N_2$ and $r_2$ 
of paired galaxies and their controls over the range of $0-1000$ kpc.
Excellent agreement between paired galaxies and their controls is seen at all separations.
\begin{figure}
\centerline{\rotatebox{0}{\resizebox{9.0cm}{!}
{\includegraphics{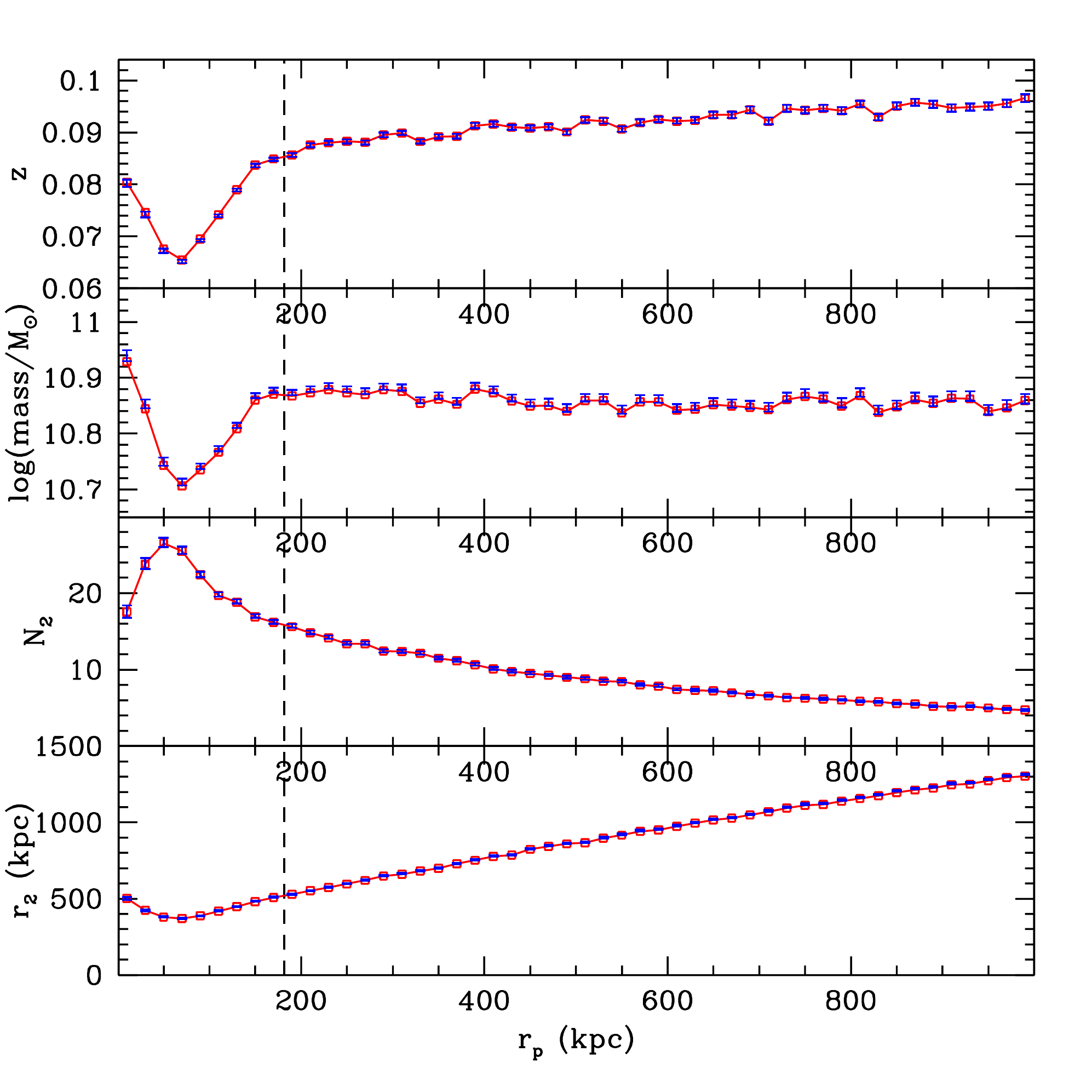}}}}
\caption{Mean redshift, stellar mass, $N_2$ and $r_2$ are plotted versus the projected separation of the 
paired galaxy's closest companion ($r_p$) 
for paired galaxies (blue symbols with error bars) and their weighted mean control galaxies (red symbols and lines).   
Error bars denote the standard error in the mean.  
For clarity of presentation, error bars are not 
shown for control galaxies; however, in every case, these error bars are smaller than those of 
the corresponding paired galaxies.  
Excellent agreement is seen between paired galaxies and their controls 
at all separations.  The vertical dashed line at 181.5 kpc identifies the separation 
within which fibre collisions are relevant (see \S~\ref{secfibre}).
\label{figcontrend}}
\end{figure}

Fig.~\ref{figcontrend} also provides us with the opportunity to explore how (and why) 
these properties vary with $r_p$.   
All four properties exhibit a smooth and monotonic dependence 
on $r_p$ over most of the range probed (at $r_p < 181.5$ kpc, 
the dependence of these properties on $r_p$ becomes more complex, due to 
spectroscopic fibre collisions; this effect is addressed below in Section~\ref{secfibre}).
Mean stellar mass exhibits very little dependence on $r_p$.   
Mean redshift increases slowly with $r_p$.  We interpret this trend as being due to the fact that 
our flux-limited galaxy sample probes further down the stellar mass function at lower redshifts, 
increasing the likelihood of finding closer companions at lower redshifts.  
$N_2$ is seen to decrease steadily with increasing $r_p$.  This trend is likely driven by several factors, 
including the fact that as local density increases, one would naturally tend to find 
additional companions at any given separation, thereby decreasing the expected 
separation of the closest companion.  Finally, $r_2$ increases steadily as $r_p$ increases.  
This behaviour is to be expected, as $r_2$ must always be greater than $r_p$ (by definition).

\section{Sources of Incompleteness}\label{seccomplete}

In Section~\ref{secsample}, we described the creation of a catalog of galaxies for which the 
closest companion of each galaxy has been identified and the local environment has been characterized.  
In Section~\ref{seccontrol}, we then outlined our approach for identifying well matched statistical 
control samples for each of these galaxies.  
Before using these data to investigate the influence of the closest companion on 
galaxy properties, we must consider various sources of incompleteness 
in these samples.  In this section, we address the effects of spectroscopic 
incompleteness, flux limits, small separations and survey boundaries on these data, 
outlining our approach to avoiding, minimizing or correcting for these sources 
of incompleteness.  

\subsection{Overall Spectroscopic Incompleteness}\label{secspeccomp}

In Section~\ref{secinput}, we described the initial selection of our redshift sample 
from SDSS, which includes restrictions on $r$-band apparent magnitudes and spectroscopic redshifts.
When comparing with the larger photometric catalogue of \citet{simard11} that this sample 
was derived from, and restricting the analysis to regions of the sky for which photometry 
and spectroscopy are both available, we estimate the overall spectroscopic completeness of 
our sample to be $\sim$ 85 per cent.  This result means that we are missing a small but significant fraction of 
galaxies which fall within the desired flux limits because they do not have reliable (if any) redshift measurements.  

This overall spectroscopic incompleteness is likely to affect our sample in a number of ways.  
First, when identifying each galaxy's closest companion, we will sometimes miss the 
true closest companion because it does not have a measured redshift.  This will 
cause us to overestimate the distance to the closest companion, and in 
some cases, it will cause a galaxy with a very close companion to be 
classified as being relatively isolated.  
Similarly, this incompleteness will sometimes cause us to underestimate $r_2$.  
When computing $N_2$, which typically lies in the range of 2-20 (see Fig.~\ref{figmethodshist}), 
we would expect to regularly underestimate $N_2$ by $\sim$ 15 percent.  
Moreover, all of these effects will degrade the underlying quality of the control sample matching, 
which depends on our closest companion identifications and environmental classifications.  

This incompleteness must be taken into consideration when using the classifications 
of any individual galaxy in our sample.  
However, since this overall spectroscopic incompleteness is quite low ($\sim 15$ per cent), 
it will not affect the identification of the closest or second closest companion for the majority of galaxies, 
and given that it will typically lead to a small reduction in $N_2$ for paired {\it and} control galaxies, 
this source of incompleteness is unlikely to have a meaningful impact on 
measurements which are averages over substantial numbers of galaxies (e.g. 
all galaxies whose companions lie in a particular range of $r_p$).  

\subsection{Spectroscopic Incompleteness Due to Fibre Collisions}\label{secfibre}

The SDSS spectroscopic sample suffers from an additional source of incompleteness that 
has much more significant implications for the study of galaxy pairs: fibre collisions.  The minimum 
physical separation between two fibres on the SDSS multi-object spectrograph leads to 
a corresponding minimum angular separation of 55 arcsec on the sky \citep{blanton03}.  
This sets a lower limit on the angular separation of a galaxy pair for which spectra can be acquired 
simultaneously for both members using a single spectroscopic plate.  At the median redshift 
of our sample (approximately 0.1), this minimum fibre separation translates to a projected 
separation of $\sim 100$ kpc.
Some regions of the sky are covered with a single plate, thereby yielding no galaxy pairs with 
separations $<$ 55 arcsec.  However, many regions are covered by two or more plates, due in 
part to planned overlap between adjacent plates.  The net result is reduced but non-zero spectroscopic 
completeness below 55 arcsec.

To illustrate the resulting spectroscopic incompleteness within our sample of paired galaxies, 
we plot redshift versus $r_p$ in Fig.~\ref{figrpz}.  The solid curved line in this 
figure depicts a fixed angular separation of 55 arcsec.  
The sharp drop in the number of detected galaxies which begins immediately to the 
left of this line is due to the minimum fibre separation.  

\begin{figure}
\centerline{\rotatebox{0}{\resizebox{9.0cm}{!}
{\includegraphics{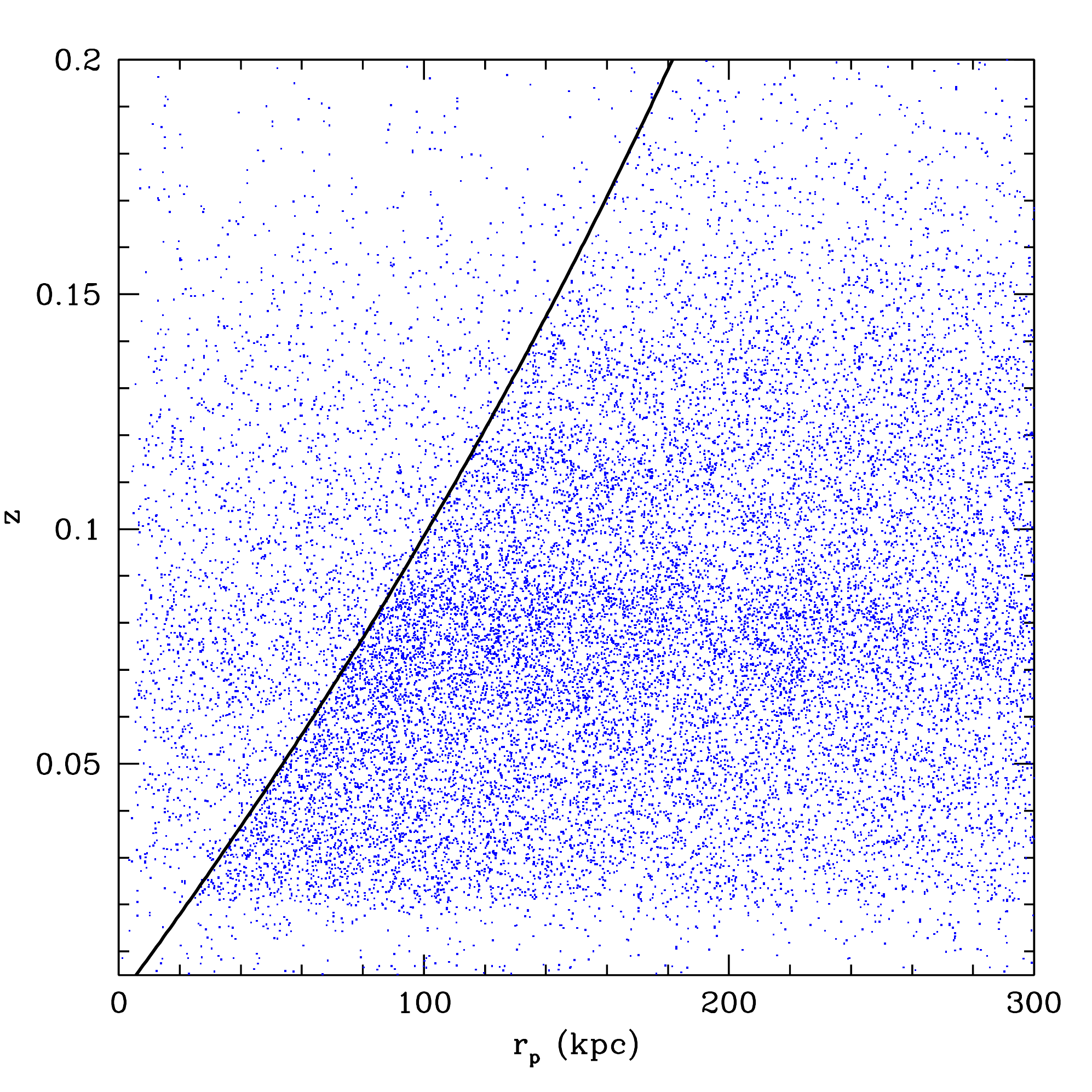}}}}
\caption{Redshift is plotted versus the projected separation of the closest companion ($r_p$) 
for 20 000 galaxies selected at random from our catalogue.
The solid curved line depicts a fixed angular separation of 55 arcsec. 
The low density of points to the left of this line is caused by 
spectroscopic incompleteness due to fibre collisions.
\label{figrpz}}
\end{figure}

Fig.~\ref{figrpz} can be used to illustrate a redshift-dependent bias that is imparted upon 
galaxy pair samples by fibre collisions.  
At large $r_p$ ($\gtrsim 180$ kpc), all pairs have separations $>$ 55 arcsec, so 
there is no spectroscopic incompleteness resulting from fibre collisions.  
As $r_p$ decreases below 180 kpc, fibre incompleteness 
biases the sample to progressively lower redshifts.  This bias is in fact clearly visible in 
the lower panel of Fig.~\ref{figcontrend}, and is seen to be strongest at $\sim 80$ kpc.
At smaller $r_p$, this bias shrinks and then disappears, since 
the closest pairs all have separations $<$ 55 arcsec (i.e., fibre collisions affect  
galaxy pairs throughout the full redshift range of our sample at these small separations).

This type of small scale spectroscopic incompleteness is a common feature of redshift surveys, 
and can been quantified and corrected for in a statistical sense.  
\citet{patton02} introduced a technique for correcting for small scale spectroscopic incompleteness 
using the ratio of spectroscopic to photometric pairs in the CNOC2 redshift survey \citep{yee00}.  
\citet{patton08} applied this methodology to the SDSS, finding a rapid drop in 
this ratio below 55 arcsec.   In several earlier papers in this series, beginning with \citet{ellison08a}, 
we addressed this incompleteness by noting that the \citet{patton08} ratio of spectroscopic to photometric pairs 
drops from about 80 per cent to 26 percent below 55 arcsec.  We then compensated for this 
factor of $\sim$ 3 reduction in spectroscopic completeness by randomly excluding 67.5 per cent of pairs 
with angular separations greater than 55 arcsec.  This approach is equivalent to randomly culling 
about two thirds of the data points to the right of the line in Fig.~\ref{figrpz}, and is successful in 
removing this obvious bias in the sample.  However, this approach also has the unfortunate consequence 
of removing the majority of the wider separation pairs.  This culling procedure has a minimal impact 
on sample size for close pairs (e.g. at $r_p < 80$ kpc), but becomes increasingly important for 
the wider pairs that are the focus of this study.

In this paper, we address this incompleteness by using statistical weights rather than culling, 
as introduced by \citet{patton13}.  Given that pairs with separations of less than 55 arcsec are 
underrepresented by a factor of $1/(1 -0.675) \sim 3.08$, we apply a fibre weight of 
$w_{\theta} = 3.08$ to every galaxy which has a closest companion within 55 arcsec.  
The success of this approach is illustrated in Fig.~\ref{figwidehist}, in which 
histograms of $r_p$, $\Delta v$ and $\mu$ are shown for all galaxies 
in our sample.  In the upper panel of Fig.~\ref{figwidehist}, the weighted and unweighted $r_p$ histograms are seen to 
increase steadily from 1000 kpc down to about 150 kpc.  However, below 150 kpc, 
the unweighted histogram turns over and then decreases at smaller separations, 
presumably as a result of fibre collisions.  However, the application of fibre weights 
removes most\footnote{The modest decrease which remains in the smallest separation bin 
is likely due to the difficulty of detecting very close companions, as discussed below in \S~\ref{secres}.}
of this apparent deficit of close pairs.  

\begin{figure}
\centerline{\rotatebox{0}{\resizebox{9.0cm}{!}
{\includegraphics{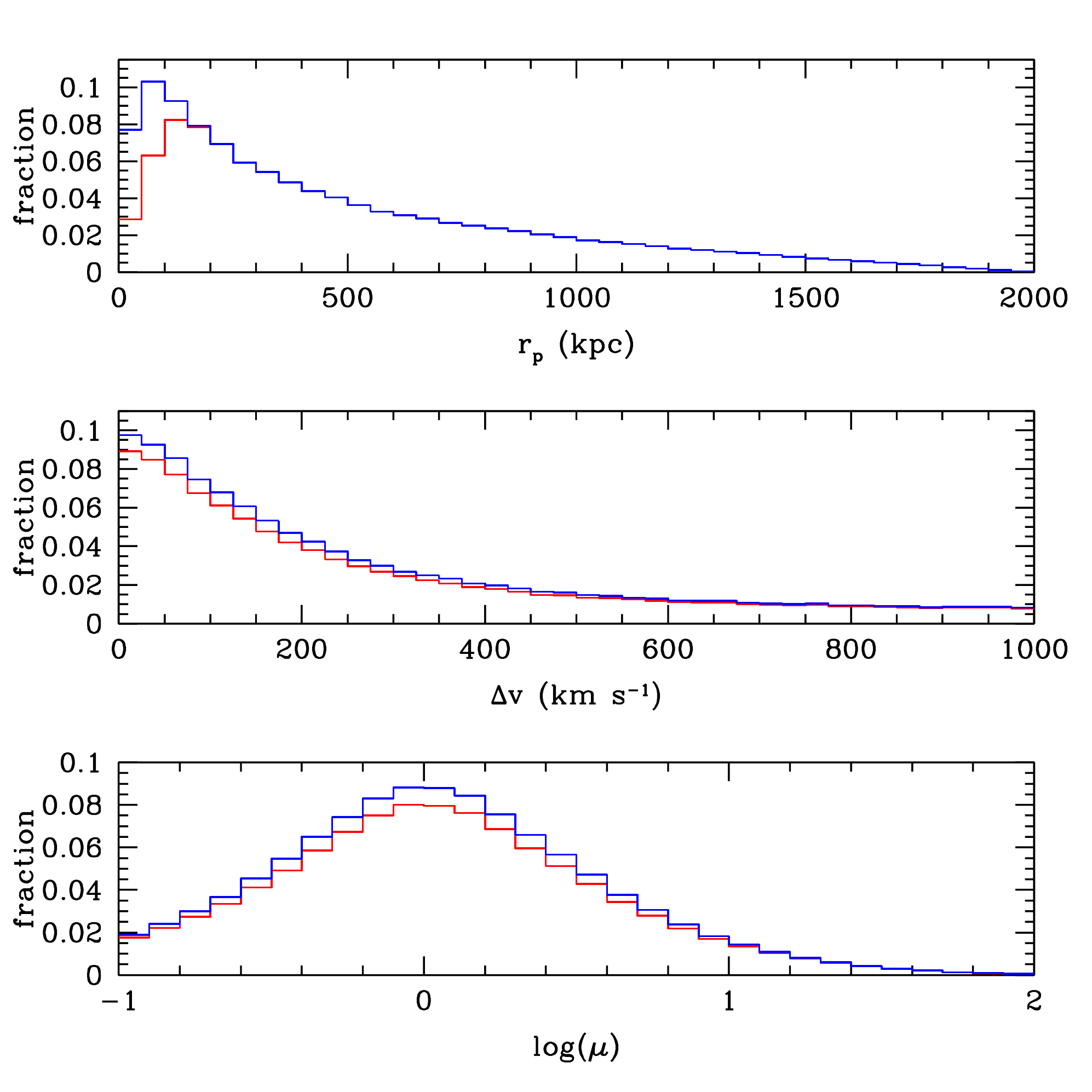}}}}
\caption{
Histograms of $r_p$, $\Delta v$ and $\mu$ are shown for 
the unweighted sample (red symbols) and the fibre-weighted sample (blue symbols).
The unweighted histograms are normalized such that the area under each is equal to one.  
The additional area under the weighted histograms is due to the application of 
fibre weights.
\label{figwidehist}}
\end{figure}

Fig.~\ref{figwidehist} also provides an overview of the properties of the closest 
companions in our sample.  While closest companions may lie anywhere in the 
ranges of $r_p < 2000$ kpc, $0 < \Delta v < 1000$~\kms~and $\mu > 0.1$, 
they are mostly likely to lie at small $r_p$, low $\Delta v$ and $\mu \sim$ 1.  
This dependence on $r_p$ and $\Delta v$ is qualitatively similar to what has been 
found in earlier close pair studies \citep{patton00,ellison10}, while the dependence on 
stellar mass ratio may instead be driven by the sample flux limits (see \S~\ref{secflux} below).

We now examine the effects of fibre weights on the dependence of redshift, stellar mass, $N_2$ and $r_2$ on $r_p$.  
In Fig.~\ref{figcontrendw}, we compare these relationships with and without fibre weights.  
The error bars on these plots refer to the standard error in the mean\footnote{For the weighted means, 
the standard error was computed using an analytic expression which has been shown to yield results 
consistent with bootstrapping \citep{gatz95}.}.
In the upper panel of Fig.~\ref{figcontrendw}, we find that the bias 
towards low redshift seen at $r_p \lesssim 180$ kpc is effectively removed by 
applying these statistical weights, yielding a smooth relationship between mean redshift and 
$r_p$ over the full range of pair separations probed (extending out to 1000 kpc in Fig.~\ref{figcontrend}), 
with the possible exception of the 
closest pairs (see \S~\ref{secres} for more on this).  Similarly, 
while the unweighted mean stellar mass shown in Fig.~\ref{figcontrendw} 
is biased towards lower stellar masses (a consequence of lower stellar mass galaxies 
being easier to detect at lower redshift), the fibre-weighted mean stellar mass is roughly independent of $r_p$ 
for all but the closest pairs.  These fibre weights also smooth out the trends in mean $N_2$ and $r_2$.  
We conclude that our fibre weights are largely successful in removing the bias due to fibre collisions. 

\begin{figure}
\centerline{\rotatebox{0}{\resizebox{9.0cm}{!}
{\includegraphics{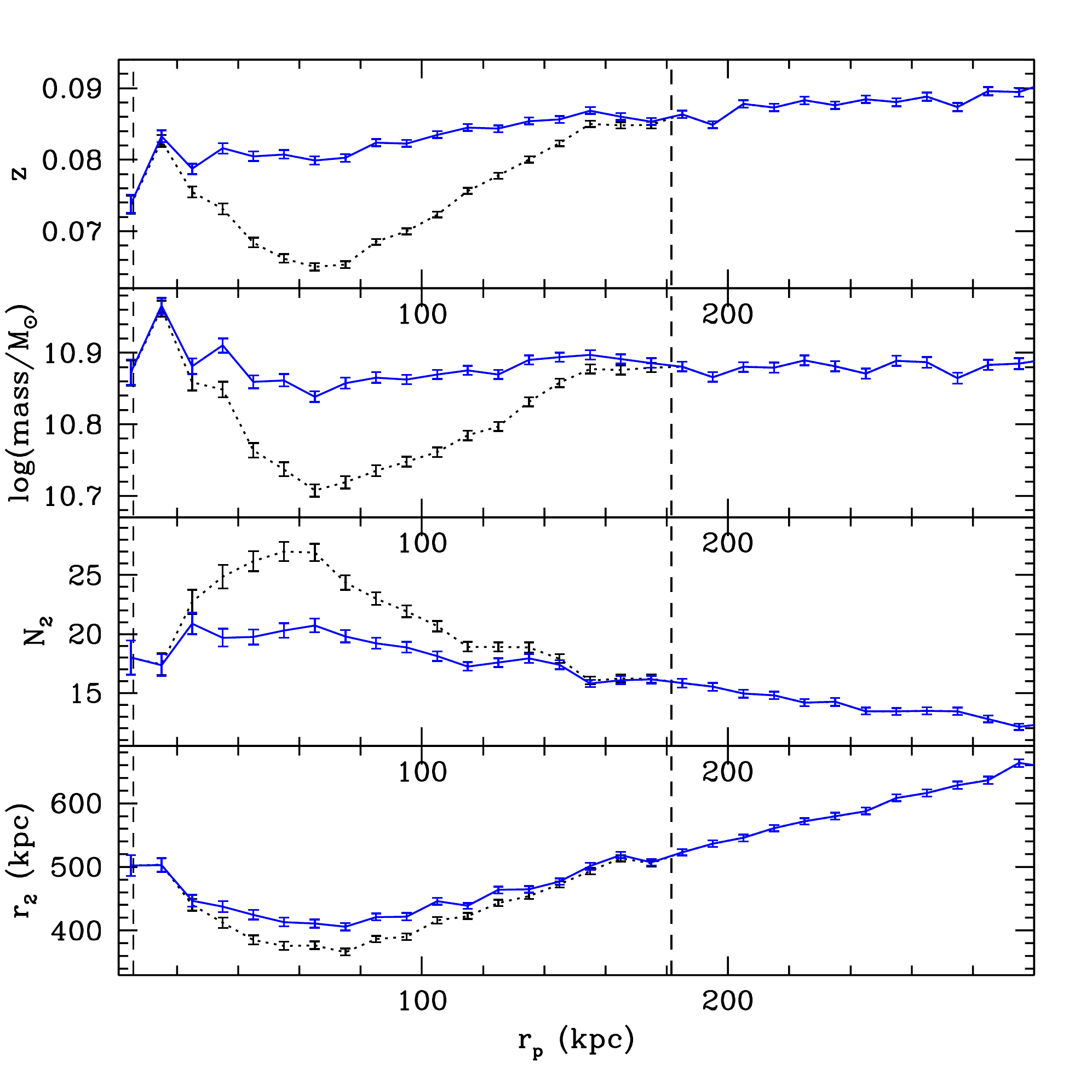}}}}
\caption{Mean redshift, stellar mass, $N_2$ and $r_2$ are plotted versus $r_p$ for paired galaxies.  
Blue symbols and solid lines denote measurements which have been corrected using fibre weights, 
whereas black symbols and dotted lines denote the corresponding unweighted measurements. 
Error bars denote the standard error in the mean.  
The vertical dashed line at 5.7 (181.5) kpc corresponds to an angular separation 
of $\theta = 55$ arcsec at the minimum (maximum) redshift of 0.005 (0.2).  
Smooth trends with $r_p$ are seen for all four properties in the weighted measurements.   
Conversely, a clear bias towards lower $z$, lower stellar mass, higher $N_2$ and lower $r_2$ 
is seen in the unweighted samples.  
\label{figcontrendw}}
\end{figure}

\subsection{Incompleteness Due to Flux Limits}\label{secflux}

When identifying each galaxy's closest and second closest companions, 
and when computing $N_2$, 
we consider all neighbouring galaxies whose stellar masses are at least 10 per cent of 
the stellar mass of the galaxy in question (see \S~\ref{secclose}).  
However, the flux limited nature of our survey ($14.0 \leq m_r \leq 17.77$; see \S~\ref{secinput}) 
means that there will often be companions with suitable stellar masses which are 
not detected.   For example, taking the crude approximation of 
stellar mass being directly proportional to $r$-band luminosity, 
the photometric depth of our sample (3.77 mag) is only 
sufficient to detect companions within a factor of 5.7 in stellar 
mass, for a galaxy in the centre of the available range in apparent magnitude. 
And if a galaxy lies near the bright (faint) flux limit, we will be strongly biased against 
finding companions which are more (less) massive than the galaxy in question.

Incompleteness due to the flux limits will affect our measurements in a similar manner to 
the spectroscopic incompleteness described in Section~\ref{secspeccomp}, 
though in a more systematic way.   The completeness will be greatest 
for companions which are similar in mass to the galaxy in question, 
and will on average become progressively worse for more unequal stellar masses.
This effect may explain why our stellar mass ratios peak at $\mu \sim 1$ (see Fig.~\ref{figwidehist}). 

The range of detectable stellar masses for companions of a given galaxy will depend 
primarily on the galaxy's stellar mass and redshift.  Fortunately, when attempting 
to detect the influence of close companions on galaxies, we compare with 
control samples which are matched on both stellar mass and redshift.  
As a result, to first order, the incompleteness due to the flux limits will 
be the same for paired galaxies and their controls.   For example, 
consider a galaxy which lies near the maximum redshift of our sample ($z = 0.2$).  
While companions with relatively low stellar masses will likely be too faint to be included 
in our flux-limited sample, leading to a relatively low value of $N_2$, the same will be true for its control galaxies, 
thereby allowing for a fair comparison between this galaxy and its controls.

\subsection{The Difficulty of Detecting Companions at Very Small Projected Separations}\label{secres}

For most galaxies in our sample, it is straightforward to discern the galaxy and its closest companion 
on SDSS imaging.  This is particularly true in the cases of the wider separation pairs which are the 
focus of the new methodology introduced in this paper.  
However, in the limit as the angular or projected physical separation between two galaxies 
becomes very small, several factors may make this task increasingly difficult.  

At the smallest separations, seeing effects may cause two galaxies to appear as a single galaxy.  
The median seeing of SDSS $r-$band images 
is about 1.5 arcsec\footnote{This corresponds to a projected separation of 2.8 kpc at the median redshift 
of our sample ($z \sim 0.1$), and a projected separation of 5.0 kpc at our maximum redshift ($z = 0.2$).}, 
with 90 per cent of the imaging being better than 1.7 arcsec \citep{dr1}.  
At somewhat larger angular separations, it may nevertheless be tricky to distinguish a galaxy from 
its closest companion if the angular separation of the pair is comparable to the angular diameter 
of one or both galaxies in the pair.  At a given $r_p$, 
these limitations will cause increasing incompleteness for pairs which are further away and/or for galaxies 
which have larger physical diameters.  In our sample, we would expect this to translate 
into a bias towards lower redshifts and smaller stellar masses at small $r_p$, 
as is seen in the innermost bin of Fig.~\ref{figcontrendw}.

Even when seeing and overlapping light profiles are not a problem, 
one may be fooled into detecting two galaxies when there is only one.  
For relatively bright galaxies, the automated SDSS deblender sometimes mistakenly identifies   
two or more galaxies, confusing sub-galactic clumps for neighbouring galaxies. 
More fundamentally, if two galaxies are in the late stages of a merger, the transition from a galaxy pair 
to a single merger remnant makes the identification of the closest companion 
inherently ambiguous.

We have avoided the worst of the deblending problems by excluding 
from our sample all galaxies brighter than $m_r = 14$.  
Our primary defence against the remaining misclassifications was to visually inspect 
every system in which the closest companion lies at $r_p < 20$ kpc, 
extending the earlier classifications of \citet{patton11}.   All probable cases of 
deblender misidentifications etc. were removed, and our sample was then regenerated, 
thereby updating the identification of the closest companion and re-measuring $N_2$, $r_2$, etc.   
These classifications greatly increase the reliability of the closest pairs in our sample.  
Of the systems that were removed by this process, the vast majority have $r_p < 10$ kpc, 
and only a handful have $r_p > 15$ kpc, indicating that 
additional inspections at larger separations are not warranted.  

Given the disparate factors which contribute to incompleteness on small scales, 
it is clearly not possible to accurately model or correct for the resulting incompleteness 
in our sample.  However, the sample that remains after visual confirmation 
can be used to identify the regime within which this incompleteness is likely to be significant.  

\begin{figure}
\centerline{\rotatebox{0}{\resizebox{9.0cm}{!}
{\includegraphics{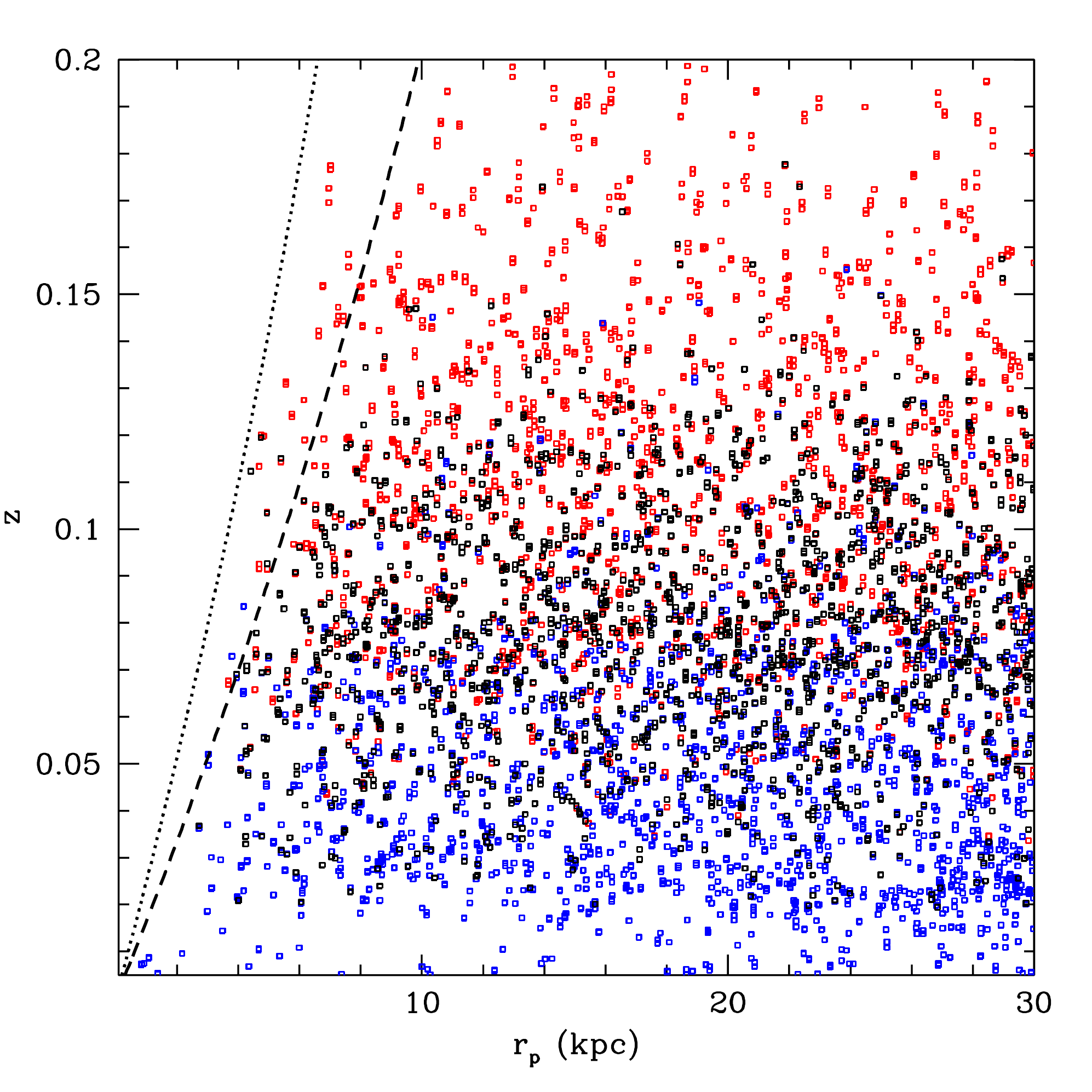}}}}
\caption{Redshift is plotted versus projected separation ($r_p$) for 
all galaxies whose closest companion lies within  30 kpc.  Galaxies are colour--coded according to 
stellar mass, with the highest mass tertile in red ($\log (M/\msun) > 10.93$), intermediate masses in black, 
and the lowest mass tertile in blue ($\log (M/\msun) < 10.36$).
The dotted line corresponds to a fixed angular separation of 2 arcsec, 
and the dashed line corresponds to 3 arcsec.  
At $r_p \gtrsim 10$ kpc, pairs are detected throughout the full 
redshift range of our sample, independent of stellar mass.  
\label{figres}}
\end{figure}

In Fig.~\ref{figres}, we plot redshift versus $r_p$ for all galaxies whose closest companion lies within 30 kpc, 
with galaxies colour-coded according to stellar mass. 
The complete absence of paired galaxies with angular separations less than 
2 arcsec (to the left of the dotted line) is broadly consistent with expectations based on SDSS seeing.  
At separations of 2--3 arcsec, some paired galaxies are detected, though there appear to 
be fewer than might be expected based on the density of points at larger $r_p$.  
Most of these pair classifications are unambiguous, as seen in the image mosaic 
of Fig.~\ref{figresmosaic}.  However, some are sufficiently close that we may be 
seeing two nuclei in a coalescing system.  
At separations $>$ 3 arcsec, Fig.~\ref{figres} indicates that we detect paired galaxies 
throughout the full redshift range of the sample.  
Moreover, since pairs of high mass galaxies close to $z \sim 0.2$ should be the hardest 
to resolve, the presence of such systems with angular separations of $\sim$ 3 arcsec 
implies that the sample as a whole is likely to be largely free of this source of incompleteness 
at and above this angular separation.  

\begin{figure}
\centerline{\rotatebox{0}{\resizebox{9.0cm}{!}
{\includegraphics{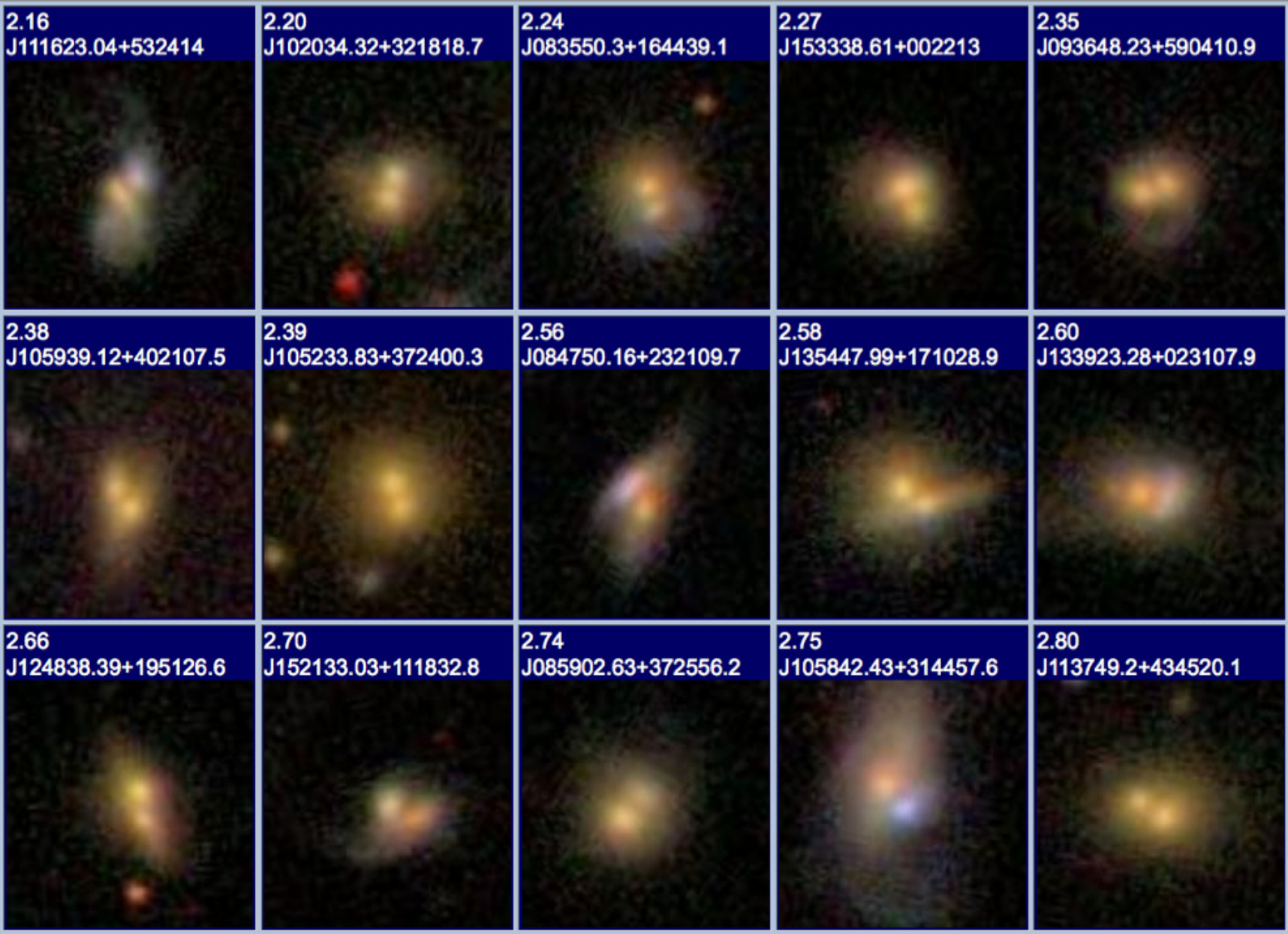}}}}
\caption{SDSS $gri$ images of the 15 closest unique galaxy pairs, sorted by angular separation.   
The angular separation (in arcsec) of each pair is labelled at the top of each image.  
In every case, two distinct galaxies can be seen.  In several cases, however, it is possible 
that we are seeing a pair that is very close to coalescence, such that a common stellar envelope 
surrounds the cores of the two precursor galaxies.
All of these galaxies lie between the dotted and dashed lines in Fig.~\ref{figres}.
\label{figresmosaic}}
\end{figure}

These qualitative findings provide guidance on potential incompleteness as a function of $r_p$.  
At $r_p > 10$ kpc, all pairs have separations $>$ 3 arcsec; therefore, it is reasonable 
to assume that our sample is unaffected by this source of incompleteness 
for any combination of redshift and stellar mass.
As $r_p$ declines below 10 kpc, an increasing fraction of paired galaxies appear to be missed, 
preferentially at higher redshift and higher stellar mass.
These trends are in fact visible in Fig.~\ref{figcontrendw}, in which a decrease in the 
mean redshift and stellar mass of the sample is seen in the smallest separation 
bin ($r_p < 10$ kpc), but not at larger separations.
Below 3 kpc, there are very few pairs, and (as one might expect) 
all of these pairs consist of relatively low mass galaxies at low redshift ($z < 0.05$).  

Finally, it is important to recognize that the preceding discussion applies only to galaxies 
in close {\it spectroscopic} pairs.  There are many more cases within SDSS where two galaxies are sufficiently close 
together that the automated algorithm identifies the system as a single galaxy.  
These misidentified galaxies are sprinkled throughout our full sample, including 
cases where galaxies have been characterized as isolated controls.
Given that the overall fraction of galaxies in 
close pairs is quite small at low redshift \citep{patton08}, the primary outcome 
is likely to be a low level of control sample contamination by potentially interacting systems.  
This base level of contamination will (to first order) be independent of 
$r_p$ (the projected separation of the nearest suitable spectroscopic companion), 
and is therefore unlikely to affect our findings in a meaningful way.

\subsection{Proximity to Survey Boundaries}\label{secbound}

The SDSS DR7 includes spectroscopic sky coverage of about 8000 deg$^2$ \citep{dr7}.  
Most of these galaxies lie in a contiguous region in the Northern Galactic Cap, 
with a smaller number of galaxies contained in three stripes in the Southern Galactic Cap.   
This uniform sky coverage ensures that most galaxies lie 
comfortably within the survey footprint.  However, when identifying each galaxy's 
closest companion (\S~\ref{secclose}), 
and especially when searching for all of its potential companions within 2 Mpc (\S~\ref{secenv}), 
we wish to ensure that the companion search radius does not overlap with the survey boundaries.
This issue is particularly important for the lowest redshift galaxies in our sample, 
since the fixed physical search radius of 2 Mpc corresponds to a relatively large 
angular search radius at lower redshift.

To deal rigorously with boundary effects, one would ideally like to use a 
detailed map of the survey geometry, beginning with all regions which were targeted 
for spectroscopy, and subsequently excluding inaccessible regions such as those 
in the vicinity of bright (especially saturated) stars or large foreground galaxies.  
This information could then be used to measure and correct for incompleteness 
due to the presence of the survey boundaries.  While information of this nature does 
exist for SDSS \citep{blanton05}, it is impractical and unnecessary to 
introduce this level of complexity into the current study.

Instead, we implement a relatively straightforward algorithm for identifying galaxies 
which are likely to lie close to the survey boundaries of our spectroscopic sample.  
We begin with the right ascension and declination of all galaxies in 
our spectroscopic sample.   For each galaxy, 
we identify all galaxies which lie within an angular separation of one degree,  
and identify the centroid of their positions.  For a uniform and isotropic distribution of neighbouring 
galaxies, this centroid will be located close to the galaxy itself.  However, 
for a galaxy located close to a survey boundary, this centroid will be significantly offset from the 
galaxy.  For example, for a galaxy lying along a straight line edge of a uniform distribution of companions, 
the centroid will be offset from the galaxy by 0.42 degrees\footnote{The centroid of a semi-circle of radius 
$r$ is located at a distance of $4r/3\pi \sim 0.42r$ from the centre of the circle.}.
In practice, the non-uniform distribution of galaxies and the complex geometry 
of the survey boundaries led us to settle on a minimum centroid offset of $\sim$ 0.3 degrees 
for identifying galaxies which lie along the survey boundaries.   
We classify all such ``boundary galaxies'' as being on or adjacent to the survey boundaries.   
The results of this approach are shown in Fig.~\ref{figbound}, 
which shows these boundary galaxies (blue symbols) enclosing the remainder of the spectroscopic sample (red symbols).
Overall, 1.0 per cent of the galaxies in our spectroscopic sample are flagged as boundary galaxies.

Having delineated the survey boundaries, we now use this information to 
find the set of galaxies which lie at least 2 Mpc away from these boundaries.  
For every galaxy in the sample, we compute the projected distance to the 
nearest boundary galaxy (hereafter $r_{\rm boundary}$).  We find that 95.3 per cent of galaxies 
in our sample lie more than 2 Mpc from all boundary galaxies, and should therefore 
have measurements of $N_2$ which are unaffected by the survey boundaries.  
We subsequently restrict our analysis to all galaxies with $r_{\rm boundary} > 2$ Mpc.

This approach appears to be effective at removing significant boundary effects from our sample.  
We note, however, that this algorithm is not precise on very small scales, since it relies on the presence of 
detected galaxies to define the edges of the distribution, rather than the actual location of 
SDSS plates on the sky.  Moreover, we do not attempt to address incomplete sky coverage which 
is due to the presence of saturated stars, bright galaxies, etc.  

Finally, having addressed the survey boundaries in the plane of the sky, 
we turn to the boundaries along the line of sight.  As stated in Section~\ref{secinput}, 
our sample is restricted to the redshift range of $0.005 < z < 0.2$.  If a galaxy 
lies near either extreme of this redshift range, some of its potential companions 
will lie outside the allowed redshift range.  In principle, this incompleteness 
could be corrected for by applying statistical weights, such as those 
introduced by \citet{patton00}.  
However, for simplicity, we instead elect to exclude from our analysis 
all galaxies which lie within 1000 \kms~ of our redshift limits, 
since this is the relative velocity threshold used when searching for potential companions (\S~\ref{secclose}).
We therefore subsequently restrict our analysis to galaxies which lie 
at $0.00836 < z  < 0.196$, while allowing their companions to lie within $0.005 < z < 0.2$.

\begin{figure}
\centerline{\rotatebox{270}{\resizebox{7.0cm}{!}
{\includegraphics{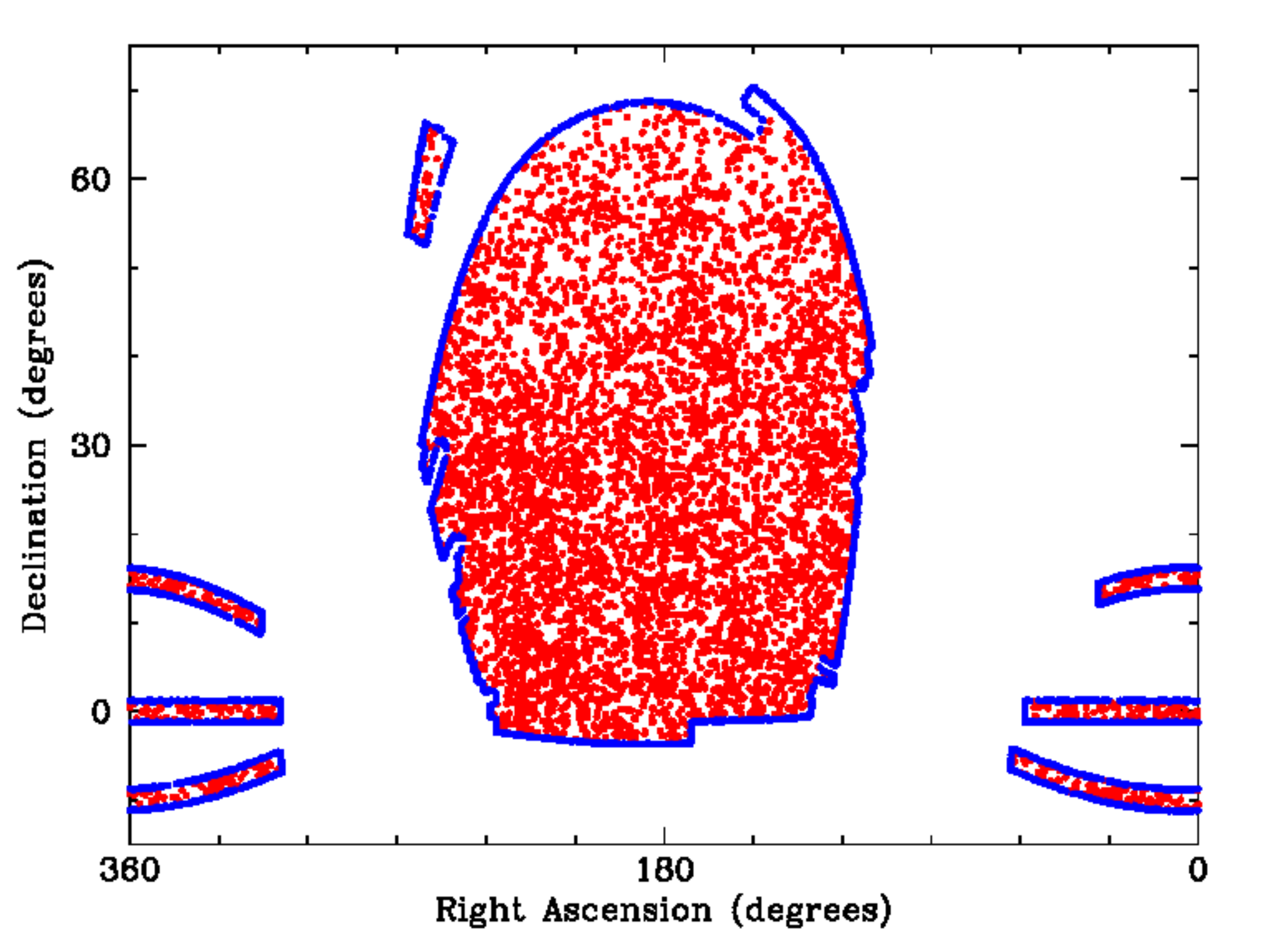}}}}
\caption{Position on the sky is plotted for all boundary galaxies in our sample (blue symbols), 
along with a random sampling of the remaining galaxies (red symbols).  
The locations of the boundary galaxies are used to estimate the projected distance to the survey boundaries 
for every galaxy in our catalogue.
\label{figbound}}
\end{figure}

\section{Application to Measurements of Galaxy Asymmetry}\label{secasym}

Having introduced our methodology for measuring the influence of the closest companion 
on galaxy properties, we now apply this approach to a set of asymmetry measurements for SDSS galaxies.
This section serves as an example of how to apply our techniques to a set of measured galaxy properties.
In addition, as we demonstrate below, the results represent a marked improvement over 
earlier efforts to examine how galaxy asymmetry is affected by the presence of close companions.

\subsection{Asymmetry Measurements}\label{secra}

The SDSS imaging of the galaxies in our sample has been processed by \citet{simard11}, 
using the {\sc GIM2D} software of \citet{simard02}.  The resulting measurements include a number 
of asymmetry-related indices.   We elect to use the $R_A$ parameter, which is defined by 
\citet{simard02}, and is based on the original definition of \citet{schade95}.  $R_A$ is a measure of 
the fraction of a galaxy's light that is left after subtracting a single component S{\'e}rsic model fit 
from the galaxy image, and then subtracting the symmetric component of the light in this residual image.  
All of the $R_A$ measurements used in this analysis were measured using the \citet{simard11} 
re-processing of the SDSS images, and were computed within two half-light radii. 
We use $R_A$, rather than the more commonly used 
parameter $R_T+R_A$, since $R_A$ should be more sensitive to tidal features, which tend  
to be asymmetric in appearance \citep{bridge10,casteels13}.  

\subsection{Sample Selection}\label{secsampleselect}

We now apply the statistical approach described earlier in this paper to these measurements 
of galaxy asymmetry.  We restrict our sample to those galaxies for which 
{\sc GIM2D} was successful in fitting a single component S{\'e}rsic model and measuring $R_A$, 
eliminating 0.2 per cent of the available galaxies.  
We note that the remaining sample spans a full range of galaxy properties, 
including both star forming and passive galaxies (unlike most earlier papers in 
this series, which were restricted to star forming galaxies).  To ensure reliable 
environmental classifications, we consider only those galaxies which have at least two 
close companions within 1.5 Mpc (see \S~\ref{secvalidation}).
We avoid boundary issues by requiring that all galaxies have $r_{\rm boundary} > 2$ Mpc 
and $0.00836 < z < 0.196$, as recommended in Section~\ref{secbound}.  
In order to focus on systems which have the potential to be undergoing significant interactions, 
we restrict our analysis to galaxies whose closest companion has $\Delta v < 300$ km~s$^{-1}$ 
and a stellar mass ratio of $0.1 < \mu < 10$ (Section~\ref{secclose}).  
Finally, in order to avoid pairs which are so close together that their overlapping light profiles 
may artificially enhance their measured asymmetries, we exclude all galaxies which have 
a companion from the \citet{simard11} sample (with or without a redshift) whose 
half-light radius (hereafter HLR) overlaps the galaxy's own HLR.
We are left with a sample of 
195 874 
galaxies which meet all of these criteria.   

For each of these galaxies, we identify a statistical control sample of at least ten galaxies, 
as outlined in Section~\ref{seccontrol}.  
The statistical weights that are applied within each control sample are determined 
by the quality of the simultaneous match in stellar mass, redshift, $N_2$ and $r_2$.  
These same weights are then applied to the measurements of $R_A$ for the control galaxies, 
yielding an estimate of the mean asymmetry for each galaxy's statistical control.  

\subsection{The Dependence of Mean Asymmetry on $r_p$}\label{secmeana}

\subsubsection{Close Pairs ($r_p <$ 100 kpc)}

In the upper panel of Fig.~\ref{figrpasymc}, 
we plot the mean $r-$band asymmetry ($R_A$) of paired galaxies (red symbols) 
and their statistical controls (grey/black symbols) as a function of $r_p$.  
In this plot, we use error bars to depict $1\sigma$ errors in the mean, 
and solid lines to depict the running mean and its $2\sigma$ uncertainty.
To compute the enhancement in asymmetry of paired galaxies, 
we divide the mean $R_A$ of paired galaxies by the mean $R_A$ of their controls, 
yielding a quotient which we denote $\mathcal{Q}(R_A)$.  
We interpret $\mathcal{Q}(R_A)$ as the enhancement in mean asymmetry due to the presence of the closest companion.  
We plot $\mathcal{Q}(R_A)$ as a function of $r_p$ in the lower panel of Fig.~\ref{figrpasymc}. 

\begin{figure}
\centerline{\rotatebox{0}{\resizebox{10cm}{!}
{\includegraphics{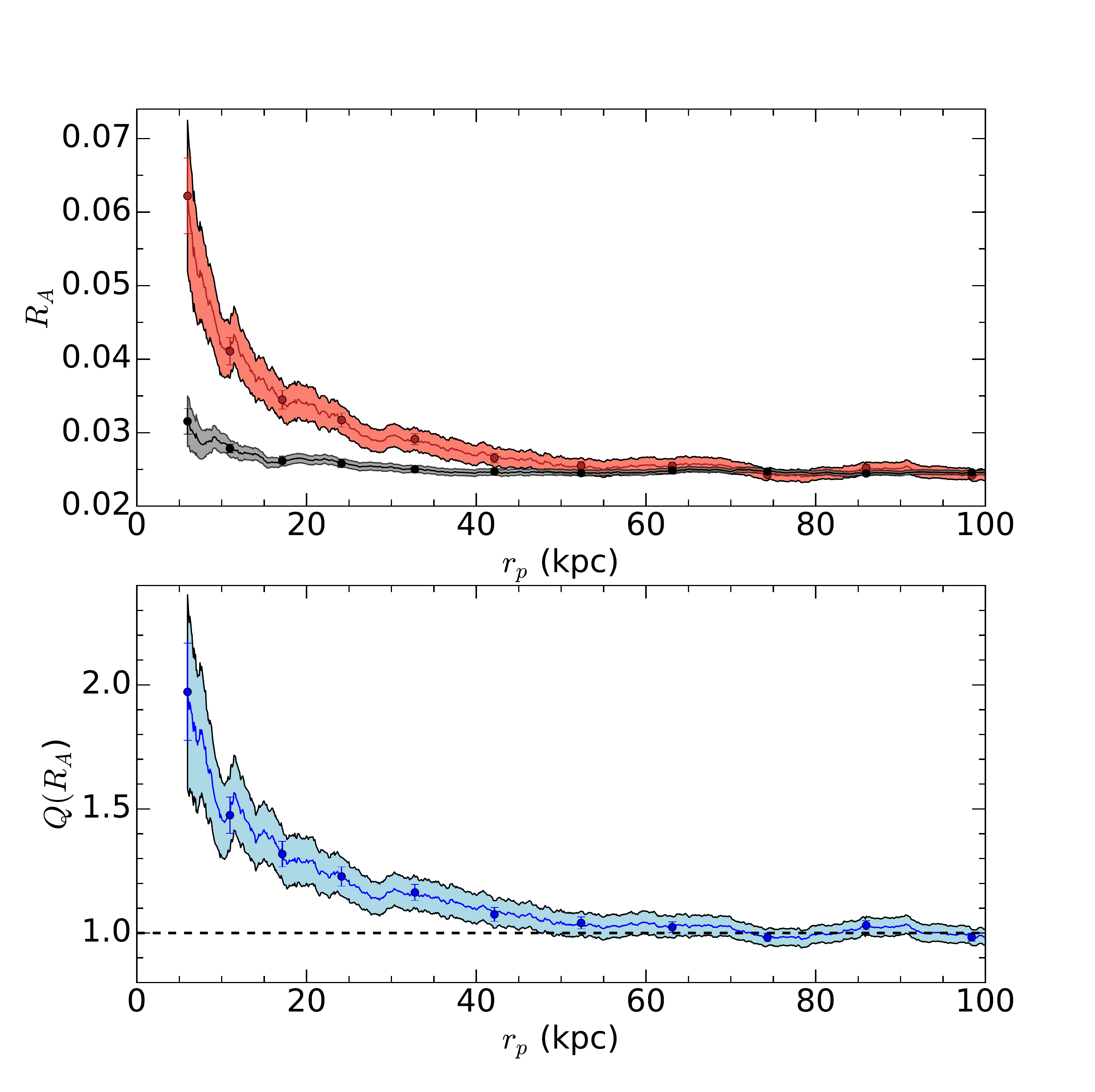}}}}
\caption{The mean $r$-band asymmetry ($R_A$) of paired galaxies (red symbols) and their controls (grey/black symbols) 
is plotted versus $r_p$ in the upper panel, while the enhancement in mean asymmetry ($\mathcal{Q}(R_A)$) 
of paired galaxies is plotted versus $r_p$ in the lower panel (blue symbols).
Filled circles with $1\sigma$ error bars depict measurements made using 
independent bins in $r_p$, with an adaptive bin width that increases from 5 kpc (at $r_p \sim 6$ kpc)
to 12 kpc (at $r_p \sim 100$ kpc).
Solid lines and shaded regions denote measurements and $2\sigma$ uncertainties 
made using rolling (not independent) bins in $r_p$.  
The horizontal dashed line in the lower panel denotes the null result of $\mathcal{Q}(R_A) = 1$ 
(no enhancement). 
\label{figrpasymc}}
\end{figure}

We find a pronounced increase in $\mathcal{Q}(R_A)$ at small separations, 
with the enhancement rising to a factor of $2.0 \pm 0.2$ at the smallest separations probed ($r_{\rm p} < 10$ kpc).   
This result is a $5\sigma$ excess above the null hypothesis of $\mathcal{Q}(R_A) = 1$.  
We detect enhancements in mean asymmetry out to $r_p = 72$ kpc (where $\mathcal{Q}(R_A)$ first drops to unity).  
This enhancement is significant at the $1\sigma$ ($2\sigma$) level at $r_p <$ 55 (47) kpc.
These findings are qualitatively consistent with predictions from merger simulations, which show that 
galaxies become disrupted during close encounters \citep{barnes92,dimatteo07,cox08,lotz08,hopkins13,patton13}, 
with morphological disturbances potentially persisting for hundreds of Myr after a strong interaction \citep{lotz10a,lotz10b}. 

Our measured enhancements in mean asymmetry 
may be compared with those of \citet{casteels14}.  
Separating their sample into six subsets according to stellar mass, they report a rise in mean asymmetry within 20-35 kpc 
for their three highest bins in stellar mass.  While they do not compare with a matched control sample, 
they find that the mean asymmetry is roughly independent of $r_p$ beyond 20-35 kpc. 
They do not report on the maximum size of the asymmetry enhancement at small $r_p$, 
but it appears to be on the order of a factor of two, 
which is comparable to the size of the maximum enhancement in mean asymmetry that we find.  

\subsubsection{Wide Pairs ($r_p >$ 100 kpc)}\label{secwide}

Figure~\ref{figrpasymc} suggests that the mean $R_A$ of paired galaxies becomes comparable to that of
their controls beyond about 70 kpc.  This convergence is consistent with the hypothesis that galaxy-galaxy interactions 
are responsible for the increased asymmetry of galaxies in close pairs.  However, with our ability to detect 
the influence of the closest companion out to much wider separations, 
we extend our analysis out to 1000 kpc in Fig.~\ref{figrpasymw}. 
We find a small but significant {\it decrease} in asymmetry ($\mathcal{Q}(R_A) < 1$) at $100 \lesssim r_p \lesssim 300$ kpc, 
reaching a minimum of $\mathcal{Q}(R_A) = 0.97 \pm 0.01$ at $r_p \sim 170$ kpc (with $5\sigma$ significance).   
At even larger separations ($r_p > 300$ kpc), $\mathcal{Q}(R_A) \sim 1$ (within $2\sigma$), although we cannot rule out 
a small deficit ($< 1$ per cent).

This small but significant decrease in the asymmetry of relatively wide pairs could be driven in part by 
earlier close encounters.  For example, the idealized merger simulations of \citet{patton13} 
include cases of galaxy pairs which have post-encounter separations of up to 220 kpc, with 
star formation having become suppressed more than $\sim$ 1.5 Gyr after the close encounter. 
It is conceivable that these galaxies might also have become less asymmetric than their 
pre-encounter progenitors, as might be expected long after 
a central burst of interaction-induced star formation has ceased.

However, dynamical arguments suggest that most pairs 
with $100 < r_p < 300$ kpc have relative velocities which are 
higher than expected for systems which have undergone previous close encounters.
A more plausible interpretation of the decreased asymmetry in these pairs is that it is driven by 
weaker and larger-scale interactions between the galaxies.  For example, processes such as 
starvation and ram-pressure stripping can quench star formation in satellite galaxies \citep{fillingham15},  
and could lead to a corresponding reduction in galaxy asymmetry.  
We defer a more detailed investigation into the nature of this decreased asymmetry to a future paper.

\begin{figure}
\centerline{\rotatebox{0}{\resizebox{10cm}{!}
{\includegraphics{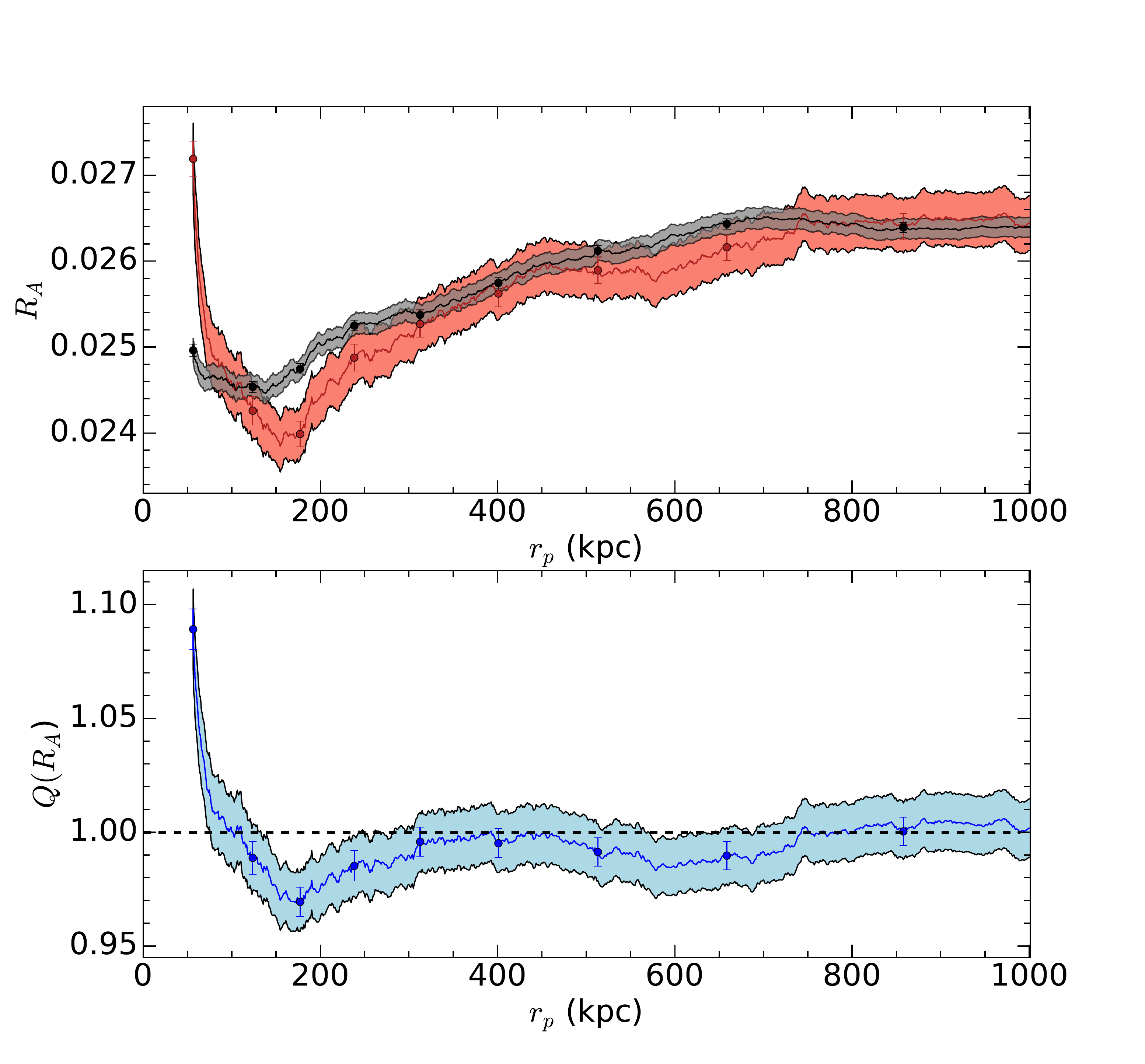}}}}
\caption{The mean $r$-band asymmetry ($R_A$) of paired galaxies (red symbols) and their controls (grey/black symbols) 
is plotted over a wide range of separations (0-1000 kpc) in the upper panel.  
The enhancement in mean asymmetry ($\mathcal{Q}(R_A)$) of paired galaxies (blue symbols) is plotted in the lower panel.
All symbols have the same meaning as in Fig.~\ref{figrpasymc}.
\label{figrpasymw}} 
\end{figure}

\subsection{Verifying the Reality of Asymmetry Enhancements Using Projected Pairs}

It is conceivable that the rise in enhancements which we have detected 
at small separations could be due in part to 
asymmetries which have been artificially enhanced due to overlapping light profiles of 
the two galaxies.   While we have attempted to minimize this effect by 
restricting our analysis to systems which are separated by 
at least one half light radius (see \S~\ref{secsampleselect}), we now test the success of 
this approach by applying our methodology to a sample of close galaxy pairs which have 
small projected separations but large relative velocities along the line of sight.  
We use the projected galaxy pairs of \citet{patton11} which have 
$3000 < \Delta v < 10 000$~\kms~and a maximum $r_p$ of 80 kpc.  
These high relative velocities ensure that the galaxies 
in these pairs cannot be interacting with one another, due to large differences in their line-of-sight distances 
from us.  As such, any detected enhancements in their asymmetries (relative to their controls) 
must be due to artificial (non-physical) enhancements in their measured asymmetries.

In the upper panel of Fig.~\ref{figrpasymp}, we plot enhancement in mean asymmetry 
as a function of $r_p$ for these projected pairs, using three different choices of minimum separation: 
zero (i.e. no minimum separation imposed), one HLR (the criterion imposed earlier) and 
two HLR (a more restrictive minimum separation).  We find no significant evidence of asymmetry 
enhancements ($\mathcal{Q}(R_A) > 1$) when using a minimum separation of one or two HLR.
However, we do find an artificial enhancement in asymmetries if no minimum separation 
is imposed (for $r_p \lesssim 25$ kpc).  This finding suggests that some of our measurements of 
$R_A$ are artificially inflated by the presence of unrelated galaxies which lie within one HLR.  
We conclude that our imposed minimum separation of 
one HLR is necessary and sufficient to ensure that our measurements of asymmetry are largely free 
of artificial enhancements due to overlapping light profiles in small separation pairs.

In the lower panel of Fig.~\ref{figrpasymp}, we directly compare the asymmetry enhancements 
of physical pairs ($\Delta v < 300$~\kms; lower panel of Fig.~\ref{figrpasymc}) 
versus projected pairs ($3000 < \Delta v < 10 000$~\kms), 
using a minimum separation of one HLR for both samples.
The absence of any dependence of $\mathcal{Q}(R_A)$ on $r_p$ in the projected pairs sample 
and the clear distinction between low velocity pairs and projected pairs at $r_p < 40$ kpc 
provide compelling support for our conclusion that close companions 
enhance galaxy asymmetries.

\begin{figure}
\centerline{\rotatebox{0}{\resizebox{10cm}{!}
{\includegraphics{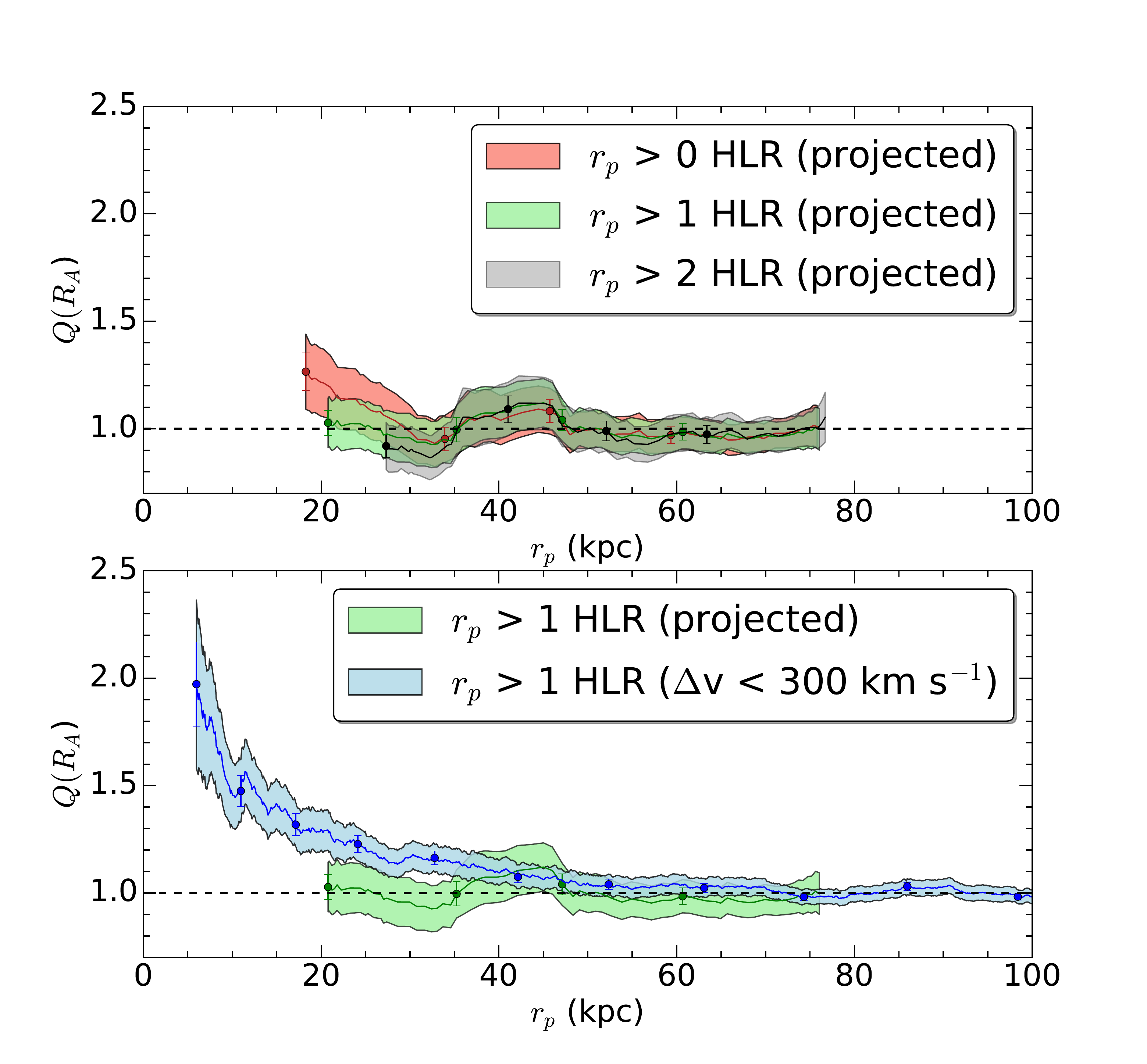}}}}
\caption{In the upper panel, the enhancement in mean asymmetry ($\mathcal{Q}(R_A)$) 
is plotted versus $r_p$ for galaxies in projected pairs ($3000 < \Delta v < 10~000$~\kms),  
using three different choices for the minimum allowed separation 
between galaxies and their closest companion (expressed in units of half-light radii).  
In the lower panel, we plot $\mathcal{Q}(R_A)$ versus $r_p$ for candidate 
interacting systems ($\Delta v < 300$~\kms; Fig.~\ref{figrpasymc}) and 
projected pairs ($3000 < \Delta v < 10~000$~\kms), using a minimum 
separation of one HLR for both.
In both panels, the horizontal dashed line denotes the null result of $\mathcal{Q}(R_A) = 1$ (no enhancement). 
All symbols have the same meaning as in Fig.~\ref{figrpasymc}.
\label{figrpasymp}}
\end{figure}

\subsection{Discussion}

We have found clear evidence for an enhancement in galaxy asymmetries that is 
due to the presence of close companions.  The size and extent of these enhancements 
is broadly consistent with the results from earlier studies \citep{depropris07,ellison10,casteels13}.  
However, the new methodology introduced in this paper has allowed us to 
confirm that these enhancements decline and effectively disappear at sufficiently wide 
pair separations, confirming a key prediction of the hypothesis that galaxy-galaxy 
interactions are responsible for the increased asymmetry in close pairs.

Our finding of significant enhancements in mean asymmetry out to separations of at least 50 kpc 
suggests that morphological signs of interactions may be shorter lived than some other 
galaxy properties.  For example, \citet{patton13} report SFR enhancements out to 
separations of about 150 kpc, using the same methodology described in this paper.  
This comparison between asymmetry and star formation rates 
is qualitatively consistent with the merger simulations of \citet{lotz08}, 
who find that enhanced star formation persists longer than morphological changes.  
However, we caution that galaxy asymmetries may in fact be enhanced 
out to larger separations than found in our study.  First, the fraction of highly asymmetric 
galaxies may be a more sensitive probe of morphological changes than 
mean asymmetry, which we have used in our study.  
In addition, the SDSS images used in our analysis are relatively shallow; deeper images 
would enable the detection of fainter and longer-lived morphogical signs of close encounters, 
as predicted by e.g. \citet{ji14}. Finally, there are likely to be more sensitive metrics of 
interaction-induced asymmetry than the $R_A$ parameter which we have used in this study 
(e.g. the shape asymmetry parameter of \citet{pawlik16}).

There are many additional questions which remain unanswered.   
Are these asymmetry enhancements driven 
by widespread low level enhancements in most relatively close galaxy pairs, 
or by strong enhancements in a small fraction of these pairs?
Are these asymmetry enhancements driven by the formation of new stars 
or by the tidal redistribution of pre-existing stars?
To what extent is {\it symmetric} residual light (e.g. bars and rings)
affected by the presence of close companions?  
What is the physical process responsible for the small but significant 
decrease in asymmetries seen at $\sim$ 100-300 kpc (\S~\ref{secwide})?
These questions are beyond the scope of this paper, but will be addressed in 
a more detailed analysis of galaxy asymmetries in a subsequent paper. 

\section{Conclusions}\label{secconclusions}

We have described a new methodology for measuring the 
influence that close companions have on galaxy properties.   
By identifying each galaxy's closest and second closest companion, 
and by comparing each galaxy to a statistical control sample which is matched on
stellar mass, redshift, local density and isolation, we are able 
to detect the influence of close companions out to arbitrarily large separations, in a wide range of environments.  
We have applied these techniques to a large sample of galaxies from the Sloan Digital Sky Survey, 
and have carefully addressed known sources of incompleteness.  

We have also demonstrated how this methodology can be applied to a 
set of measured galaxy properties, by analyzing the mean asymmetry of 
galaxies as a function of pair separation.  We find that close companions 
enhance mean galaxy asymmetry out to separations of at least 50 kpc, 
with the enhancement in mean asymmetry rising to a factor of $2.0 \pm 0.2$ at 
projected separations $<$ 10 kpc.
We find no evidence for enhanced asymmetries in close 
projected pairs ($\Delta v > 3000$~\kms), thereby confirming that 
the enhanced asymmetries are not an artifact of overlapping light profiles in close galaxy pairs.
These results are consistent with the interpretation that the detected enhancement in 
the asymmetries of close pairs is due to galaxy-galaxy interactions.
We also find a small ($< 3$ per cent) but 
significant (up to $5\sigma$) deficit in asymmetry at wider separations 
($\sim$ 100-300 kpc) which may be driven by larger scale 
interactions between the galaxies rather than close interactions.

Our methodology can be used to explore how a wide range of galaxy properties 
are influenced by the presence of a close companion.   For example, 
in several earlier papers in this series, we have detected differences 
in metallicities \citep{scudder12}, colours \citep{patton11} and AGN fractions \citep{ellison11} 
out to the 80 kpc limit of our sample of SDSS close pairs.   
The sample outlined in this paper could be used to determine how much further out 
these differences are found.  In addition, with our measurements of stellar mass, local density 
and isolation, it would be possible to examine how these interaction-induced changes 
depend on stellar mass, stellar mass ratio and environment.  
Finally, the techniques introduced in this paper are well suited to 
the study of galaxies in cosmological simulations, enabling a direct comparison 
between observations and simulations of interacting galaxies.

\section*{Acknowledgements}

DRP and SLE gratefully acknowledge the receipt of NSERC Discovery Grants which
helped to fund this research.  PT acknowledges support from NASA ATP Grant NNX14AH35G.
JM is supported by NSF grant AST--1516364.

The SDSS is managed by the Astrophysical Research Consortium for the 
Participating Institutions. The Participating Institutions are the 
American Museum of Natural History, Astrophysical Institute Potsdam, 
University of Basel, University of Cambridge, Case Western Reserve University, 
University of Chicago, Drexel University, Fermilab, the Institute for 
Advanced Study, the Japan Participation Group, Johns Hopkins University, 
the Joint Institute for Nuclear Astrophysics, the Kavli Institute for 
Particle Astrophysics and Cosmology, the Korean Scientist Group, 
the Chinese Academy of Sciences (LAMOST), Los Alamos National Laboratory, 
the Max--Planck--Institute for Astronomy (MPIA), the Max--Planck--Institute for 
Astrophysics (MPA), New Mexico State University, Ohio State University, 
University of Pittsburgh, University of Portsmouth, Princeton University, 
the United States Naval Observatory, and the University of Washington. 





\bsp	
\label{lastpage}
\end{document}